\documentclass[showkeys,showpacs,longbibliography,aps,prb,twocolumn]{revtex4-2}
\usepackage{amsmath}
\usepackage{color}
\usepackage{hyperref}
\usepackage{graphicx}
\usepackage{setspace}
\usepackage{tikz}

\newcommand{\VEC}[1]{{\boldsymbol{ #1}}}

\newcommand{\Fig}{{Fig.}}
\newcommand{\be}{\begin{equation}} \newcommand{\ee}{\end{equation}}
\newcommand{\bea}{\begin{eqnarray}} \newcommand{\eea}{\end{eqnarray}}

\newcommand{\sbs}{Sb$_2$S$_3$}
\newcommand{\mos}{MoS$_2$}
\newcommand{\bise}{Bi$_2$Se$_3$}
\newcommand{\half}{\frac{1}{2}}

\begin{document}
\title{
Efficacious symmetry-adapted atomic displacement method for lattice dynamical studies
}

\author{Chee Kwan Gan}
\email{ganck@ihpc.a-star.edu.sg}
\affiliation{Institute of High Performance Computing, 1 Fusionopolis Way, \#16-16 Connexis 138632, Singapore}
\author{Yun Liu}
\affiliation{Cavendish Laboratory, University of Cambridge, 19 JJ Thomson Avenue, Cambridge CB3 0HE, United Kingdom}
\author{Tze Chien Sum}
\affiliation{Division of Physics and Applied Physics, School of Physical and Mathematical Sciences, Nanyang Technological University, 21 Nanyang Link 637371, Singapore}
\author{Kedar Hippalgaonkar}
\affiliation{Institute of Materials Research and Engineering (IMRE), A*STAR Agency for Science, Technology and Research, 2 Fusionopolis Way,
\#08-03 Innovis 138634, Singapore}
\affiliation{School of Materials Science and Engineering, Nanyang Technological University,  639798, Singapore}

\date{Sep 17, 2020}

\begin{abstract}

Small displacement methods have been successfully used to calculate the
lattice dynamical properties of crystals.  It involves displacing atoms
by a small amount in order to calculate the induced forces on all atoms
in a supercell for the computation of force constants.  Even though
these methods are widely in use, to our knowledge, there
is no systematic discussion of optimal displacement directions from
the crystal's symmetry point of view nor a rigorous error analysis of
such methods.  Based on the group theory and point group symmetry of a
crystal, we propose displacement directions, with an equivalent concept
of the group of $k$, deduced directly in the
Cartesian coordinates rather than the usual fractional coordinates,
that maintain the theoretical maximum for the triple product $V$
spanned by the three displacements to avoid possible severe roundoff
errors. The proposed
displacement directions are generated from a minimal set
of irreducible atomic displacements
that keep the required independent force calculations to a minimum.
We find the error in the calculated force constants explicitly depends on
the inverse of $V$ and inaccuracy of the forces. Test systems such as Si,
graphene, and orthorhombic \sbs{} are used to illustrate the
method.  Our symmetry-adapted atomic displacement method is shown to be very robust in treating
low-symmetry cells with a large `aspect ratio' due to huge differences
in lattice parameters, use of a large vacuum height, 
or a very oblique unit cell due to unconventional choice of primitive lattice vectors.
It is expected that our atomic displacement
strategy can be used to address higher-order interatomic interactions
to achieve good accuracy and efficiency.
\end{abstract}


\maketitle

\section{Introduction}

Lattice dynamical studies\cite{Born56-book} are important for an 
understanding of the 
phase stability,\cite{VanDeWalle02v74,Gan10v49} ferroelectric transition,\cite{Zhong94v72} Raman and 
infra-red spectroscopies,\cite{Zhao13v13,Chong14v90,Giovanni18v5} 
phonon-mediated superconductivity,\cite{Giustino17v89}
ferroelastic transition,\cite{Togo08v78} 
and
thermodynamics of materials.\cite{Grimvall1999-book,Mujica03v75}
Even though the concept of phonon emerges as a result of a harmonic approximation, its simple extensions
via the quasiharmonic approximation (QHA)\cite{Mounet05v71}
and Gr\"uneisen formalism\cite{Allen20v34,Malica20v32} 
allow thermal properties due to anharmonic effects such as 
lattice thermal conductivities\cite{Toher14v90}, a key quantity that determine the figure of merits for thermoelectrics\cite{Snyder08v7,Madsen16v213,Gorai17v2}, as well as 
thermal expansion  coefficients\cite{Gan15v92,Arnaud16v93,Gan16v94,Romao17v96,Gan18v151,Gan19v31} to be evaluated.  An attempt has been made to extract  the third order interatomic force constants from
the standard phonon calculations through the evaluation of Gr\"uneisen parameters\cite{Lee17v96}.
Phonon databases are being built\cite{Petretto18v5,TogoPhononDB2020-link} with the
help of high-throughput frameworks\cite{Curtarolo13v12,Ong13v68,Pizzi16v111} for data mining and machine learning.

The methods to calculate phonon frequencies
and eigenvectors naturally fall into two distinct approaches. 
One approach is based on the displacement methods\cite{Parlinski97v78,Wang14v185,Wang16v2,Ackland97v9,Togo15v108,Togo2020-github,Alfe09v180,Kresse95v32,Gan10v49} that gain popularity due to their
simplicity in implementations. In these methods,
 the force constants can be deduced from the induced forces via
the Hellman-Feynman theorem\cite{Feynman39v56} when 
a small displacement on an atom in an otherwise perfect supercell is made.
The possibility to calculate exact 
phonon frequencies at commensurate $\VEC{q}$ vectors
using smaller non-diagonal unit cells\cite{Lloyd-Williams15v92} paves the way for
very practical applications of these methods.
The second 
approach is
based on the density-functional perturbation theory.\cite{Baroni01v73,Gonze97v55a}
In this approach, the methods are very versatile because
small unit cells rather than huge supercells
are required. Very accurate energy derivatives could be analytically calculated
within a computer code\cite{Giannozzi09v21,Gonze09v180}.
Both approaches could be used together to 
complement each other\cite{Fu19v100} for the extraction of higher-order 
interatomic interactions.\cite{Parlinski18v98}

A somewhat simple and effective implementation of
a small displacement method is to use
a few pre-selected directions in the fractional coordinates that are to be acted upon by
appropriate space group operations to deduce the displacement
directions as long as the volume $V$
spanned by the actual displacement directions is nonzero. 
This strategy may have been inspired by the fact that
space group operations\cite{ITtable06-book} usually act on the 
positions of atoms in fractional coordinates within a unit cell.
However, as we shall show later, such
implementation may result in a nonoptimal choice of the displacements of atoms that 
affects the accuracy of the force constants and eventually the lattice dynamical
properties. We use a generic
 orthorhombic system of space group $Pnma$ with lattice parameters 
$a$, $b$, and $c$
to show that the volume $V$ may deviate from its ideal value of 1 and 
scale unfavorably
as $2(a/b)^{-1} \rightarrow 0 $ as  $a \gg b$. 
In the case of simulating a graphene sheet using a supercell method, it is 
shown that a large vacuum thickness could reduce the ideal $V=1$ to 
a value that scales as $(c/a)^{-2}$ where $c$ is the vacuum height and $a$ the 
hexagonal lattice parameter for graphene. In this paper
we propose a displacement method that can be applied to any crystal, encompassing the 
entire 32 crystallographic point groups and 230 space groups.
The displacement directions are deduced directly in the Cartesian
coordinates rather than fractional coordinates
that are designed to maintain (i) the theoretical maximum for the 
triple product $V$ spanned by the three displacements
to avoid possible severe roundoff errors, and (ii) a minimal set of irreducible atomic displacements for independent force calculations.
To achieve these aims we rely on the concepts of the star of $k$ and the group of $k$\cite{Dresselhaus2008-book}, defined originally in the 
reciprocal space but extended to real space in this paper.
Various test systems such as Si, graphene, and orthorhombic
\sbs{} are used to illustrate the method.
This paper is organized as follows. In Section~\ref{sec:method}
we provide the full details 
of our displacement method to make judicious atomic displacements for all crystal symmetries.   Section~\ref{sec:FC_error} 
presents an error analysis for the force constants. Results are shown in Section~\ref{sec:results}. Finally we conclude in Section~\ref{sec:conclusion}. Appendix
A illustrates more clearly how $V$ may scale poorly with the lattice parameters or the choice of
oblique unit cell for a few selected cases.

\section{Methodology}
\label{sec:method}
First we define a matrix 
\be
A=[ \VEC{a}_1|\VEC{a}_2 | \VEC{a}_3] 
\label{eq:A}
\ee
where
the $i$th column of $A$ is taken from 
the basic lattice translation vector $\VEC{a}_i$ of a crystal.
For simplicity
we use $A$ to describe a primitive cell but it could be easily 
extended to deal with a conventional unit cell or even a nonconventional 
unit cell.
A space group operator $\{ R | \VEC{t} \}$ 
(in the Seitz notation) corresponds
to a rotation matrix 
\be
R_c = A^{-1} R A
\label{eq:Rc}
\ee
in the Cartesian coordinates, which is restricted to 
either $p$ or ${\overline p}$ for $p= 1$, $2$, $3$, $4$, and $6$. 
$p$ means a $p$-fold rotation in the international 
notation, while $ {\overline p}$ means an
improper rotation (i.e., a $p$-fold rotation followed by an inversion).
The 230 space groups are built on top of 
the 32 crystallographic 
point groups (see e.g., Refs.~[\onlinecite{Dresselhaus2008-book,Burns85-book}]).

In the small displacement method\cite{Liu14v16} for
a phonon calculation of a crystal,
if the intrinsic symmetries of a 
space group are not
utilized, we have to sequentially
displace all $N_1$ atoms in a primitive cell
embedded in a large $n_1 \times n_2 \times n_3$ supercell of
$n_1 n_2 n_3 N_1$ atoms
along the $x$, $y$, and $z$ Cartesian axes in 
both positive and negative directions,
resulting in
$6N_1$ different 
supercells, 
each
with one atom displaced slightly compared to the unperturbed supercell (we call this the all-displacement method).
Each of the $6N_1$ supercells will be treated independently
where the induced forces are to be calculated so that
the interatomic force constants can be deduced.
Due to a computational cubic scaling with respect to the number of atoms, most density-functional theory (DFT)
implementations\cite{Payne92v64,Gan01v134} face a severe practical
issue to handle $6N_1$ supercells to evaluate the induced forces. It is therefore important
to reduce the number of calculations as much as possible.

We now review how the force constant matrix $\Psi_{ij}$
 between
the $i$th atom in the primitive cell and the $j$th atom in the
supercell may be calculated.
This is achieved by sequentially displacing the $i$th
atom in the primitive cell by three displacement vectors $\lambda \VEC{d}_k^i$ in the 
Cartesian coordinates, $k = 1, 2, 3$, where
$\VEC{d}_k^i$ is a unit vector. 
$\lambda$ is the magnitude of the displacement which is typically
$0.010 \sim 0.015$~\AA.
For each displacement vector $\lambda \VEC{d}_k^i$,
we calculate the induced forces
on all atoms in the supercell, in particular the induced force
$\VEC{F}_k^j$ on the $j$th atom, $k=1,2,3$.
If we make $\VEC{F}_k^j$ to 
constitute the $k$th column of a matrix $F^j$ where
\be
F^j = [\VEC{F}_1^j | \VEC{F}_2^j | \VEC{F}_3^j ]
\ee
and similarly we
let $\VEC{d}_k^i$ to constitute the $k$th column of a displacement matrix $d^i$ where
\be
d^i = [\VEC{d}_1^i | \VEC{d}_2^i | \VEC{d}_3^i] 
\label{eq:dmat}
\ee
then the required force constant $\Phi_{ij}$ can be calculated from
\be
F^j = \lambda \Phi_{ij} d^i
\label{eq:Phi}
\ee
From Eq.~\ref{eq:Phi} 
we see that the displacement vectors $\VEC{d}_k^i$ do not
need to point along the conventional Cartesian axes but in
any directions as long as $d^i$ is not singular.

A first level of a possible reduction of the number of calculations can be made
if an inequivalent atom in the primitive cell, 
where its position is commonly known 
as the Wyckoff position,
could be mapped under space group operations
to its equivalent atoms in the primitive cell, resulting in a 
`star of $k$', which is understood to be a set of atoms.
We note that the concept of the star of $k$ is originally defined in 
the reciprocal space\cite{Dresselhaus2008-book}.
When handling `a star of $k$', we may arbitrarily choose
any atom in a set to be a representative atom (the so-called
inequivalent atom) for the 
purpose of generating the rest of equivalent atoms in the same set. 
The $3\times 3$ force constant matrix $\Phi_{ij}$
between the $i$th inequivalent atom in the
primitive cell and another $j$th atom in the supercell
can be ``copied'' out\cite{Liu14v16} to 
another $3\times 3$ force constant matrix $\Phi_{i'j'}$ between 
the $i'$th equivalent atom and the $j'$th atom using
\begin{equation}
\Phi_{i'j'} =  R_c \Phi_{ij} R_c^T
\end{equation}
$R_c$ is as defined in Eq.~\ref{eq:Rc} and $ \{ R | \VEC{t} \}  $
maps the $i$th atom to the $i'$th atom, and
the $j$th atom to the $j'$th atom.  $R_c^T$ is the matrix transpose of $R_c$.
With this strategy alone (we call this the 6-displacement method), 
if there are $N_0$ inequivalent atoms in a primitive cell, 
then only $6 N_0$ calculations are needed. 

However, it may still be possible to reduce the 6 calculations for each
of $N_0$ inequivalent atoms using the 
site symmetry\cite{Burns85-book} of an inequivalent atom. The
site symmetry (this has the equivalent concept of the group of $k$, see
for example Ref.~[\onlinecite{Altmann-book}]) of an inequivalent
atom is one of the 32 
crystallographic point groups that
leaves the position of an inequivalent atom invariant in periodic
sense.  Note that two different inequivalent atoms from two different 
stars of $k$ may not have the same site symmetries.

Now we shall discuss how the site symmetry can be used
to reduce the number of displacements for an inequivalent atom.
Suppose a displacement $\VEC{d}_1^i$ has been applied to an inequivalent
$i$th atom and the forces on all atoms in the supercell have been found.
If an element $\{R|\VEC{t}\}$ in the site symmetry of the $i$th
atom is applied to the supercell with the $i$th atom that has been displaced by
$\VEC{d}_1^i$, then the net effect of the operation
is to rotate the original displacement
$\VEC{d}_1^i$ to become a displacement $\VEC{d}_2^i = R_c \VEC{d}_1^i$
on the $i$th atom.  The operation $\{R|\VEC{t}\}$ also reshuffles
the positions of all atoms in the supercell, as well as to rotate the
induced forces caused by $\VEC{d}_1^i$ on all atoms in the supercell. This
crucial observation implies 
that without doing an independent (probably expensive)
induced force calculation
due to $\VEC{d}_2^i$, we are able to just use the information due to
$\VEC{d}_1^i$ to give us all force information for
a virtual $\VEC{d}_2^i$ displacement.  
If there is yet another operation $\{R|\VEC{t}\}$ that could
rotate $\VEC{d}_1^i$ to $\VEC{d}_3^i$, then again an independent displacement
of $\VEC{d}_3^i$ does not need to be carried out.  However, in the case when
there is no extra operation in a site symmetry that can
generate an independent $\VEC{d}_2^i$ or $\VEC{d}_3^i$,
then we have no choice
but to carry out necessary separate displacements and find the induced
forces to fill up the necessary force field for the $\Phi_{ij}$ calculation.

In an elegant implementation, the displacement in the Cartesian
coordinates $\VEC{d}^i_k$ can be generated from the directions defined
in the fractional coordinates $\VEC{g}_k^i$ where 
$ \VEC{d}^i_k =  \frac{A \VEC{g}_k^i}{ |  A \VEC{g}_k^i  |   }$.  
$\VEC{g}_{k}^i$ may be chosen from a set of $S$ 
that consists of \emph{nonzero}
vectors of the form $ (e_1, e_2, e_3)^T$, where $e_n = 0, \pm
1$ for $n= 1,2,3$ for simplicity.  By systematically applying the elements
of the site symmetry to vectors in $S$, one may find a minimal set
$S_i$ that contains between one to three vectors. 
$S_i$ will generate three $\VEC{g}_k^i$ that form a nonzero determinant for $d^i$.
This implementation has a slight drawback that may be illustrated by
two similar
orthorhombic systems Bi$_2$S$_3$\cite{Zhao11v84} and 
\sbs\cite{Liu14v16,Chong14v90,Gan15v92}
in the $Pnma$ setting, where $a \sim 11.3$~\AA, $b \sim 3.8 $~\AA, and $c \sim
11.1$~\AA. Here $a$ is about three times larger than $b$.  
There are twenty atoms in the primitive cell, with
five inequivalent atoms on the $4c$ Wyckoff site. 
The site symmetry is a group of mirror reflection (a two-element group).
The reflection operator $\overline{2}$
reflects the system
across the $xz$ plane and maps a direction in 
the fractional coordinate 
$ \VEC{g}_1^i = (1,-1,0)^T $ to $ \VEC{g}_2^i = (1,1,0)^T$.
However, these two directions in the fractional coordinates correspond to
$(a,-b,0)^T$ and $(a,b,0)^T$ in the Cartesian coordinates which
subtend an angle not equal to an ideal
angle of $90^\circ$ since $a \neq b$.
The angle between the two displacements in
the Cartesian coordinates is actually
given by $2\tan^{-1}{\frac{b}{a}}$ (see Appendix A for a detailed discussion).
Hence if $a$ is much larger than $b$, the two vectors $\VEC{d}_1^i$
and $\VEC{d}_2^i$ are nearly parallel to each other in the Cartesian coordinates.
However the operation is a physical reflection in the $xz$ plane and
hence it is possible to force the angle between the two displaced vectors
to be exactly $90^\circ$, thereby achieving the largest 
determinant for $d^i$ of $1$ for matrix inversion in Eq.~\ref{eq:Phi}.

In another example, a graphene sheet of
a lattice constant $a$ and a vacuum thickness $c$ may use $\VEC{g}_1^i =
(1,0,1)^T$ to generate $\VEC{g}_2^i = (0, 1, -1)^T$ and $\VEC{g}_3^i = 
(-1,-1,1)^T$ with the point group operations. 
This means only one independent force calculation is to be performed.
However, an analysis shows that $V = \det d^i = \frac{ a^2 c \sqrt3  }{ 2(a^2 + c^2)^{3/2}   }$, which 
goes to $\frac{\sqrt3/2}{(c/a)^2}$
for a large $c/a$ ratio. With\cite{Gan10v81} $c=20$~\AA{} and $a = 2.471$~\AA{}, 
$V = 0.013$. However, with the method to be developed later,
we find that the angles between any two displacements taken from $\VEC{d}_k^i$, $k=1,2,3$
can be made $90^\circ$ thereby making $V$ achieves its largest 
value of 1.

To develop a sense of how the inaccuracy of forces and
 $V = \det d^i$ may affect the
accuracy of force constants, we consider the force constants in the
$xy$ plane for the $4c$ site of the
$Pnma$ space group. Here $\VEC{d}_1^i = \frac{1}{\sqrt{a^2 + b^2}}(a,-b)^T$ 
and $\VEC{d}_2^i = \frac{1}{\sqrt{a^2+ b^2}}(a,b)^T$.
From Eq.~\ref{eq:Phi}, we have
\be
[\VEC{F}_1^j | \VEC{F}_2^j ] = \lambda [\Phi_{ij}^1| \Phi_{ij}^2] 
\frac{1}{\sqrt{a^2 + b^2}}
  \begin{pmatrix}
       a &  a \\
       -b &  b \\
  \end{pmatrix}
\ee

We then have
\be
\Phi_{ij}^2 = \frac{1}{2\lambda} \left( \frac{a^2}{b^2}  +1 \right)^{\frac{1}{2}}
 (\VEC{F}_2^j -  \VEC{F}_1^j)
\ee
If $a \gg b$, then
$\VEC{d}_1^i$ and $\VEC{d}_2^i$ are almost parallel to each other resulting in
very similar forces $\VEC{F}_1^j$ and $\VEC{F}_2^j$.
If the force $\VEC{F}_1^j$ due to $\VEC{d}_1^i$ is not 
determined accurate enough, $\VEC{F}_2^j$ will inherit the same inaccuracy since it 
is `copied' from $\VEC{F}_1^j$ through a symmetry operation, then
$\Phi_{ij}^2$ will be inaccurate due to a large roundoff error that is amplified by 
a large geometry factor $ \left( \frac{a^2}{b^2}  +1 \right)^{\frac{1}{2}}$.

Now we present a method
that will systematically deduce the displacement directions 
directly in the \emph{Cartesian coordinates} for a
forward difference scheme with the aim of 
maintaining a largest possible magnitude for the determinant of $d^i$ and
a minimum number of independent displacements.
The displacement method
must also maintain the minimal number of independent displacements when a 
central difference scheme is used for an improved accuracy for $\Phi_{ij}$.
For a central difference scheme, we need to displace the atoms
in $-\VEC{d}_k^i$, $k = 1, 2, 3$ where $\VEC{d}_k^i$ have been chosen 
for a forward difference scheme. We note that two operations
are able to map $\VEC{d}_k^i$ to $-\VEC{d}_k^i$: one is the inversion
operator, and the other a 
2-fold rotation.

\begin{table}
\begin{center}
\begin{tabular}{lllll}
\hline
Crystal & Symmetry  & $n_{\rm FD}$ & $n_{\rm CD}$ & \\
\hline
\hline
Triclinic &    $C_1$  & 3 & 6 \\
 &    $C_i$ & 3 & 3 \\
\hline
Monoclinic &    $C_2$ &  2 & 3 \\
 &    $C_{s}$ & 2  & 4 \\
 &    $C_{2h}$ & 2 & 2 \\
\hline
Orthorhombic &    $D_2$ & 1 & 2 & \\
 &    $C_{2v}$ & 1 & 2 & \\
 &    $D_{2h}$ & 1 & 1 & \\
\hline
Tetragonal &    $C_4$ & 1 & 2 & \\
 &    $S_4$ & 1 & 2 & \\
 &    $C_{4h}$ & 1 & 1 & \\
 &    $D_4$ & 1 & 1 & \\
 &    $C_{4v}$ & 1 & 2 & \\
 &    $D_{2d}$ & 1 & 1 & \\
 &    $D_{4h}$ & 1 & 1 & \\
\hline
Trigonal &    $C_3$ &1 & 2 \\
 &    $S_6$ & 1 & 1 & \\
 &    $D_3$ & 1 & 1 & \\
 &    $C_{3v}$ & 1 & 2 & \\
 &    $D_{3d}$ & 1 & 1 & \\
\hline
Hexagonal  &    $C_{6}$ & 1 & 2 & \\
 &    $C_{3h}$ & 1 & 2 & \\
 &    $C_{6h}$ & 1 & 1 & \\
 &    $D_{6}$ & 1 & 1 & \\
 &    $C_{6v}$ & 1 & 2 & \\
 &    $D_{3h}$ & 1 & 1 & \\
 &    $D_{6h}$ & 1 & 1 & \\
\hline
Cubic &    $T$ & 1 & 1 & \\
 &    $T_h$ & 1 & 1 & \\
 &    $O$ & 1 & 1 & \\
 &    $T_d$ & 1 & 1 & \\
 &    $O_h$ & 1 & 1 & \\
\hline
\end{tabular}
\end{center}
\caption{
$n_{\rm CD}$ ($n_{\rm FD}$) is the minimal number of displacements per atom for a central (forward) difference scheme.
}
\label{tab:32PGs}
\end{table}

The task may at first seem arduous for all 32 crystallographic 
point groups (see Table~\ref{tab:32PGs}). However, there are actually 
four distinct cases to consider.\footnote{See an implementation of our algorithm in fm-forces.f90
  from https://github.com/qphonon/atomic-displacement}
The first case deals with
the triclinic crystals and covers
point groups 1 to 2 ($C_1$ and $C_i$)
that involve a single $1$ or ${\overline 1}$ operation.  
The second case handles the monoclinic crystals and covers
point groups 3 to 5 ($C_2$, $C_s$, $C_{2h}$) 
that involve a single $2$  or ${\overline 2}$ operation. 
The third case deals with the orthorhombic 
crystals and covers point groups 6 to 8  ($D_2$, $C_{2v}$, $D_{2h}$).
These point groups possess
three operations involving 
either $2 $ or ${\overline 2}$ that are perpendicular
to one another.
Finally the fourth case
deals with the remaining four crystal systems, i.e., 
tetragonal, trigonal, hexagonal, and cubic crystals and covers
point groups 9 to 32.
These point groups have a single $3$, ${\overline 3}$, $4$, 
or ${\overline 4}$ operation.

Usually, the principal axis (the axis with the highest symmetry) 
is not pointing along the $z$ axis but
along a vector $\VEC{n}$ characterized by 
the polar angle $\theta$ and azimuthal angle $\phi$.
To ease the discussion of the determination
of $\VEC{d}_k^i$, we 
transform all operations $R_c$ to $R_c'$  in the site symmetry according to
$R_c' = Q^{-1} R_c Q$, $Q = R_z(\phi) R_y(\theta)$. Here
$R_y (\theta) $ [$R_z (\phi)$]
is a rotation of angle $\theta$ ($\phi$) around the $y$ ($z$) axis.
Under such transformation, the operation with the 
principal axis $R_{\VEC{n}}(\eta)$  will be mapped to
$R_z(\eta)$ where the rotation axis is now along the $z$ axis. 
This can be easily seen since $R_{\VEC{n}}(\eta)$ is transformed to
$ R_z(\eta) $ via $R_{z}(\eta) = Q^{-1} R_{\VEC{n}}(\eta) Q$ or
$ R_{\VEC{n}}(\eta) = Q R_{z}(\eta)  Q^{-1}$.
Once we find $\VEC{d'}_k^i$ in the rotated frame,
we obtain $\VEC{d}_k^i = Q \VEC{d'}^{i}_k$.

\subsection{The first case}
First we consider the triclinic cell where there are two point groups $C_1 (1)$ and $C_i ({\overline 1})$.
For $C_1 (1)$ the symmetry is so low that
we simply propose to displace the $i$th atom in $x+$, $y+$, and $z+$, resulting in three independent displacements for 
a forward difference scheme. 
In this case, $V = \det d^i$ attains its largest possible value of unity.
For a central difference scheme, we need to displace the atom 
in $x-$, $y-$, and $z-$, resulting
a total of six displacements.
For $ C_i ({\overline 1})$, an
inversion operator exists that could map $\VEC{d'}_k^i$ to $-\VEC{d'}_k^i$ 
hence we need to do the same number of displacements (i.e., $3$)
for
both the forward and central difference schemes (see the third
 row of Table~\ref{tab:32PGs}).

\subsection{The second case}
For the monoclinic cell there are three point groups to consider.  We first
consider $C_2 (2)$, we propose to use a
first displacement $\VEC{d'}^i_1 =
(0,-1,1)^T$. Under a 2-fold rotation around the $z$ axis,
we generate a second displacement
$\VEC{d'}_2^i = (0, 1,1)^T$ from $\VEC{d'}_1^i$.
We exhaust all operations in this point group,
hence we propose a second independent displacement vector
$\VEC{d'}_3^i = (1,0,0)^T$.  
This choice makes $V$ to attain its largest possible value of unity.
For a central difference scheme, we need
to displace the $i$th atom in $-\VEC{d'}_1^i$ independently
since there is no operation in the site symmetry 
that is able to map $\VEC{d'}_1^i$ to its negative. 
However, the 2-fold rotation
is able to map $\VEC{d'}_3^i = (1,0,0)^T$ to its negative $(-1,0,0)^T$. 
Hence the total number of
displacements is three for the point group $C_2(2)$ instead of four for a central difference scheme.  
For the second point group $C_{s} (m)$. There is no 
operation that can map $\VEC{d'}_1^i$ and $\VEC{d'}_3^i$
to their negatives. Hence the number of independent displacements is
four for a central difference scheme. 
The third point group to consider is $C_{2h}$. Since there is an inversion operator
the number of displacements
for the forward and central difference schemes are the same, which is two.

\subsection{The third case}
For the next three point groups for the orthorhombic cells $D_2 (222)$,
$C_{2v} (mm2)$, and $D_{2h}(mmm)$, we simply need to displace an atom
in $\VEC{d'}_1^i = \frac{1}{\sqrt{3}} (1,1,1)$ direction.  This direction is obtained by
maximizing a test vector $(t,u,v)^T$ and its images $(t,-u,-v)$ (under a
$2$-fold rotation around the $x$-axis) and $(-t,u,-v)$ (under a $2$-fold
rotation around the $y$-axis) thus forming a triple product of $ V = 4tuv$. The magnitude of the displacement vector
is constrained according to
$t^2+ u^2 +v^2 = 1$. A standard Lagrange multiplier method 
gives a possible solution 
$\VEC{d'}_1^i = \frac{1}{\sqrt{3}} (1,1,1)^T$
that delivers the largest $V = \frac{4}{\sqrt{27}} = 0.7698$.

For $D_2$ and $C_{2v}$ there is no
operation that could map $\VEC{d'}_1^i$ to $-\VEC{d'}_1^i$ 
hence we need to do
a total of two
displacements for a central difference scheme.  
However, $D_{2h}$ has an inversion operator
that results in the same number (i.e., one)
of independent displacement
for both the forward and central difference schemes.

\subsection{The fourth case}
For the next 24 point groups from $C_4 (4)$ to $O_h (m{\overline 3}m)$
covering tetragonal, trigonal, hexagonal, and cubic cells, we note
that we have either $3$ or ${\overline 3}$ or $4$ or ${\overline 4}$ that
is able to generate $\VEC{d'}_2^i$ and $\VEC{d'}_3^i$ from just one
$\VEC{d'}_1^i$.
For these point groups, using the Lagrange multiplier method, we find
that a single displacement is able to give a maximum $V$ of
$\frac{4}{\sqrt{27}}$ and $1$ for $C_4$ and $C_3$, respectively,  with a
test vector of $\VEC{d'}_1^i = (t,u,\frac{1}{\sqrt{3}})^T$, $t^2 + u^2
= \frac{2}{3}$.  In order to prepare $ \VEC{d'}_1^i $ to be mapped under
an existing $C_2$ operation (which is unfortunately 
missing in eight point groups $C_4$, $S_4$, $C_{4v}$, $C_3$, 
$C_{3v}$, $C_6$, $C_{3h}$, and $C_{6v}$)
 to $ - \VEC{d'}_1^i $
with the $C_2$ rotation axis pointing along $(n_x, n_y, n_z)$
direction, we can find the values of $t$ and $u$
by solving the simultaneous equations 
$t^2 + u^2 = \frac{2}{3}$ and $n_x t + n_y u + \frac{n_z}{\sqrt{3}} = 0$.
This is equivalent to solving a quadratic equation $(n_x^2 +
n_y^2) u^2 + \frac{2}{\sqrt{3}} n_y n_z u + \frac{1}{3} \left( n_z^2 -
2n_x^2\right) = 0$. 
To see more clearly the nature of solutions to the quadratic equation,
without loss of generality we assume the axis of rotation for
$C_2$ is in the $yz$ plane making an angle 
$\theta$ with the $z$ axis where $(n_x, n_y, n_z) = 
(0,\sin\theta,\cos\theta)$. This gives
two solutions $(t,u,v) = 
(\pm \sqrt{1- \frac{1}{3\sin^2\theta}}, 
- \frac{\cos\theta}{\sqrt{3} \sin \theta}, \sqrt{\frac{1}{3}})$.
For $\theta = \frac{\pi}{2}$, $(t,u,v) = 
( \pm \sqrt{ \frac{2}{3}  }, 0, \sqrt{\frac{1}{3}})$.  
As $\theta$ is decreased from $\frac{\pi}{2}$, 
two solutions will approach one another along the circumference
of a circle and coincide with one another 
when $\theta $ becomes a critical
angle $\theta_c  = \sin^{-1} \sqrt{\frac{1}{3}} = 35.26^\circ$ 
and the solution is $(t,u,v) = ( 0, -\sqrt{\frac{2}{3}}, \sqrt{\frac{1}{3}}   )$.
When  $\theta < \theta_c$, there is no real solution.

\section{Error analysis for the force constants}
\label{sec:FC_error}
Now we perform a detailed 
error analysis for the displacement method. We shall 
focus on $\Phi_{ij}$ which is
the $3\times 3$ force constant matrix block between
the $i$th atom and the $j$th atom
based on Eq.~\ref{eq:Phi}.
Since atom pairs are now understood, we 
suppress the
index $i$ in $d^i$ and index $j$ in $F^j$  for  forces in Eq.~\ref{eq:Phi}.
We let the exact force $\VEC{F}_k$
to differ from the approximate force $\VEC{f}_k$
by an error term $\lambda \VEC{\epsilon}_k$ that
measures the intrinsic inaccuracies where
\be
\VEC{f}_k = \VEC{F}_k +  \lambda\VEC{\epsilon}_k, \ \ k = 1, 2, 3
\ee

The calculated force constant block with $d$ 
is given by
\bea
\Phi_d &=& 
[ \VEC{f}_1 | \VEC{f}_2 | \VEC{f}_3] (\lambda d)^{-1} \nonumber 
\\ &=& F(\lambda d)^{-1} + [\VEC{\epsilon}_1 | \VEC{\epsilon}_2 | \VEC{\epsilon}_3] 
d^{-1} \nonumber
\\ &= & \Phi_e +  \epsilon d^{-1}
\label{eq:Phid}
\eea
where $\Phi_e$ is the exact force constant matrix and
$\epsilon = [\VEC{\epsilon}_1 | \VEC{\epsilon}_2 | \VEC{\epsilon}_3]$.
Eq.~\ref{eq:Phid}
clearly shows that the error can be attributed by the 
intrinsic inaccuracy of the force and 
the inverse of $d$, which could potentially be
large if $d$ is near singular.
In the absence of the exact force constant $\Phi_e$ to calculate errors
we rely on
\be
\phi_d  - \phi_{d_0} = \epsilon ( d^{-1} - d_0^{-1})
\label{eq:Err}
\ee
where $d_0$ is chosen to be with the largest possible $V$.
We will study the inaccuracy of 
the calculated force constant as a function of $V$
in the next section.

\section{Results}
\label{sec:results}
To confirm the methodology outline above, we 
carry out density-functional theory (DFT) calculations within the local density
approximation, with projector augmented-wave (PAW) pseudopotential
as implemented in the Vienna Ab initio Simulation Package (VASP)
\cite{Kresse96v6} on a generic system of silicon. 
We use a 2-atom primitive rhombohedral cell
and a $3\times 3\times 3$ supercell for the phonon calculations.
The cutoff energy for the plane-wave basis set is
307~eV. We use a $3\times 3\times 3$ 
$k$ mesh for the supercell calculations. 
The magnitude of the displacement $\lambda $ is $0.015$~\AA.
Because of the presence of a $C_3$ rotation with its rotation axis
along the $(1,1,1)^T$ direction in the Cartesian
coordinates and another $C_2$ 
operation in Si just one displacement is
sufficient for a central difference scheme for the phonon calculations. 
However, since we want to study the effect of $V$ on the 
phonon dispersions, we will still utilize $C_3$ in
the $(1,1,1)^T$ direction to rotate $\VEC{d}_1 = (x,y,y)^T$
that results in just one displacement
if the forward difference scheme is used. However another
displacement in the negative 
direction is needed for a central difference scheme
since the $C_2$ operation is now no longer able to map 
$\VEC{d}_1$ to $-\VEC{d}_1$.
Under $C_3$ with its rotation axis in $(1,1,1)^T$, we have
$\VEC{d}_2 = (y, x,y)$, and $ \VEC{d}_3 = (y,y,x)$. This gives
$V = x^3 - 3xy^2 +2y^3$. With the constraint on the
magnitude of $\VEC{d}_1$, which gives $2x^2 + y^2 = 1$,
we solve for
$x$ and $y$ using a Newton-Raphson scheme for a predetermined $V$. We notice that 
$V \rightarrow 0$ as
$x$ and $y$ go to $\frac{1}{\sqrt{3}}$. 

\begin{figure}
\centering\includegraphics[width=9.2cm,clip]{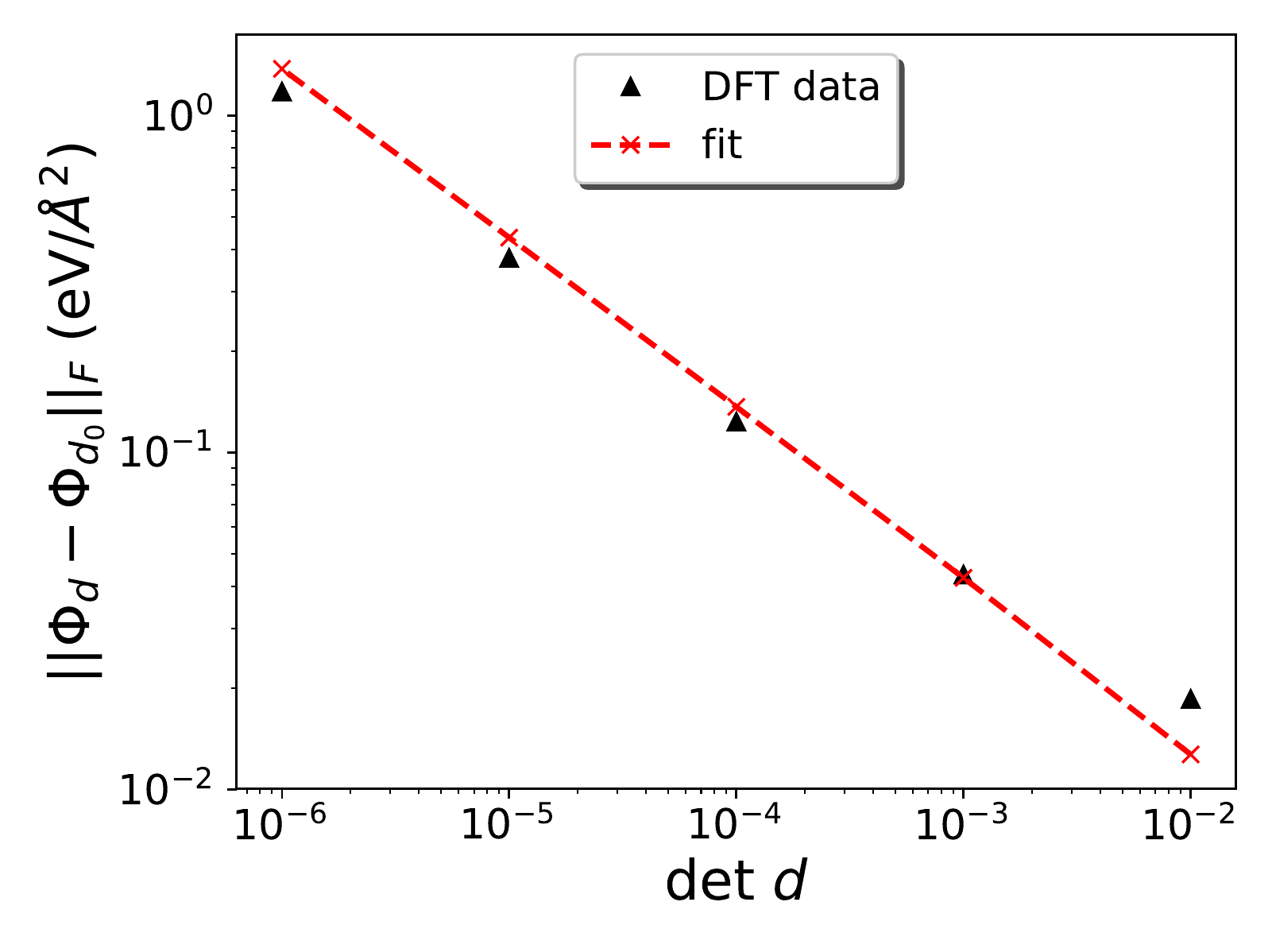}
\caption{
The error (filled triangle) characterized
by the Frobenius norm of the force constant block $\phi_d  - \phi_{d_0}$ in Eq.~\ref{eq:Err}
as a function $V = \det d$. The fitting form according to the RHS of Eq.~\ref{eq:Err} gives
$\epsilon_0 = 3.7\times 10^{-4}$~eV/\AA$^2$.
}
\label{fig:2020-04-22-mpl-kphonon-det-gives-error-fit.pdf}
\end{figure}

We investigate the error of the force constants
as measured by the Frobenius norm  $||\phi_d - \phi_{d_0}||_F $
as a function $V$ using Eq.~\ref{eq:Err}. We use $\det d_0 = 1$. 
\Fig~\ref{fig:2020-04-22-mpl-kphonon-det-gives-error-fit.pdf}
shows the error of the force constants
linearly increases with decreasing $V = \det d$ .

\begin{figure}
\centering\includegraphics[width=9.2cm,clip]{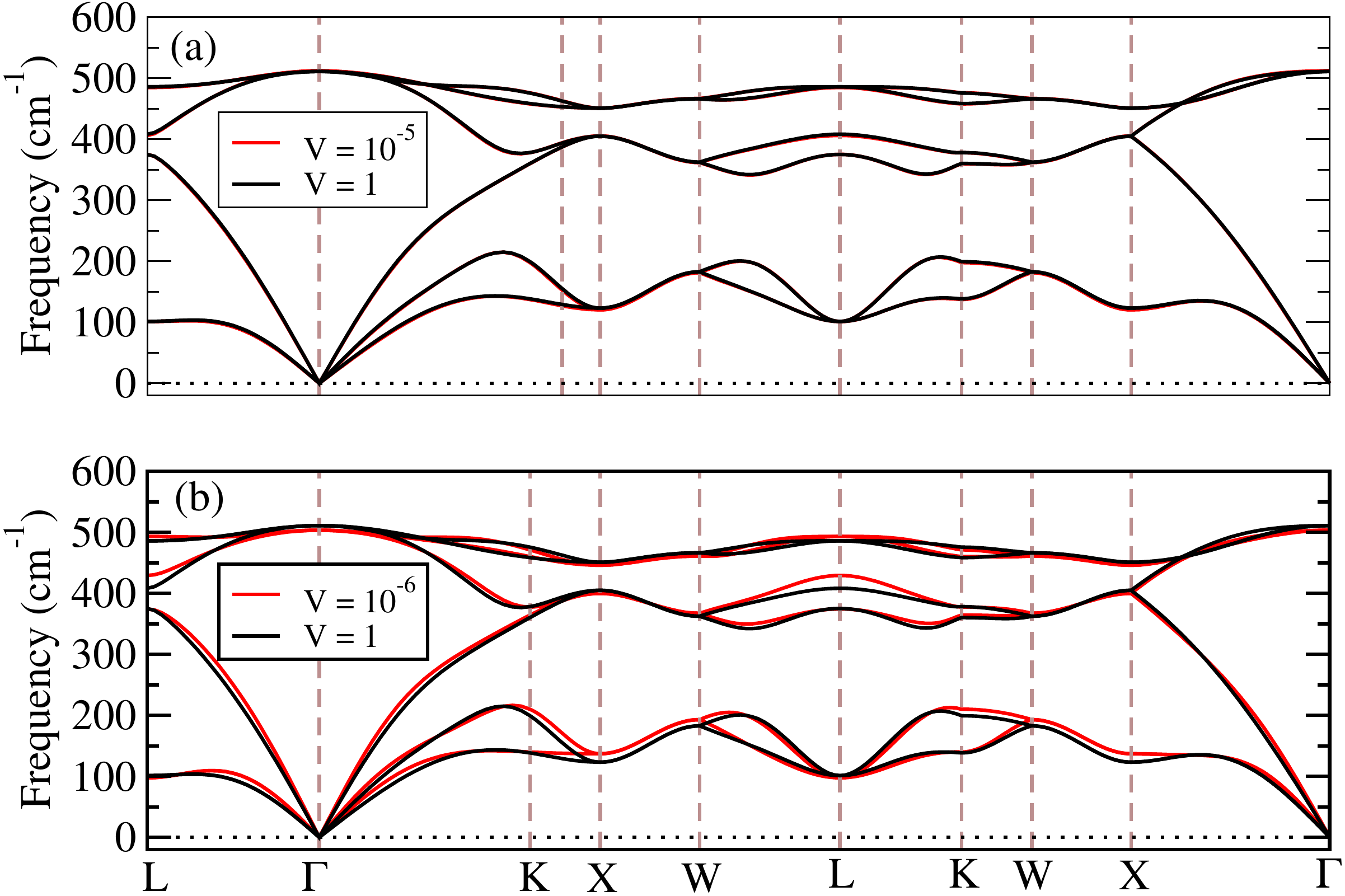}
\caption{
(a) The comparison of phonon dispersions obtained with $V = 10^{-5}$ and $V=1$. 
(b) The comparison of phonon dispersions obtained with $V = 10^{-6}$ and $V=1$.
}
\label{fig:2020-04-24-Si-Phonon-Det-1d-5-and-Det-1d-6.pdf}
\end{figure}

For an estimate of $\epsilon $ in Eq.~\ref{eq:Err} without a full
knowledge of the error of the induced forces, we proceed
by assuming $\epsilon= \epsilon_0 J$, which
is characterized by a single
error parameter $\epsilon_0$ and an appropriate $J$ matrix. Our 
aim is to 
estimate  $\epsilon_0$ that measures the inaccuracy of the force by
fitting the RHS of Eq.~\ref{eq:Err}.
We observe that
in the limit when $V $ is very small, the three displacement vectors
$\VEC{d}_1$, $\VEC{d}_2$, and $\VEC{d}_3$ become closer and closer 
to one another and converge to
$\frac{1}{\sqrt{3}}(1,1,1)^T$. 
To maximize error it is therefore reasonable to assume
$\VEC{\epsilon}_1 = \varepsilon_0 (1,-1,-1)^T$, 
$\VEC{\epsilon}_2 = \epsilon_0 (-1,1,-1)^T$, and $
\VEC{\epsilon}_3 = \epsilon_0 (-1,-1,1)^T$. Under such assumption and 
with Eq~\ref{eq:Err} we obtain a rather good fit and deduce $\epsilon_0 = 3.7\times 10^{-4} $~eV/\AA$^2$.
This in turn gives a reasonable estimate of the force inaccuracy of $\lambda \epsilon_0 = 5.6\times 10^{-6}$~eV/\AA.
\Fig~\ref{fig:2020-04-24-Si-Phonon-Det-1d-5-and-Det-1d-6.pdf} shows  phonon dispersions
with different $V$ values. 
Noticeable differences in phonon dispersions are observed when $V $ reaches $10^{-6}$.

\begin{figure}
\centering\includegraphics[width=9.2cm,clip]{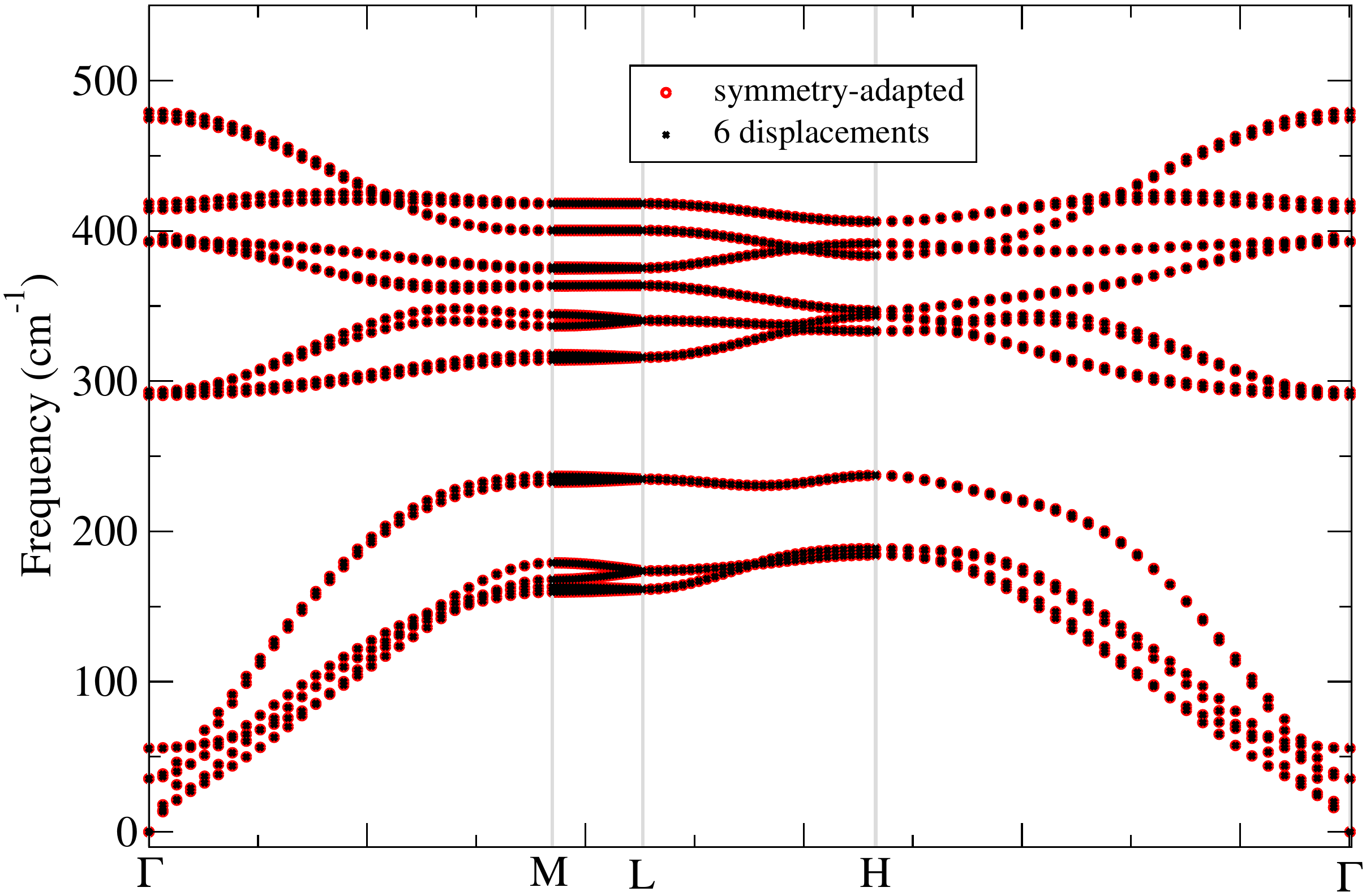}
\caption{
Phonon dispersions of hexagonal \mos{} with $a=3.123$ and $c = 12.087$~\AA{} obtained with 
(a) symmetry-adapted atomic displacements and (b) six displacements for each of the two
inequivalent atoms.
The effect of longitudinal optical (LO) and transverse optical (TO) splitting\cite{Liu14v16} has been taken account.
A $3\times 3 \times 2$ supercell is used. The cutoff energy for the plane-wave basis set
is $700$~eV. 
A $k$ mesh of $2 \times 2 \times 1$ is 
used for electronic relaxation. 
The selected $\VEC{q}$ points (in $\VEC{b}_1$, $\VEC{b}_2$, and $\VEC{b}_3$) are 
$\Gamma=[0,0,0]$, $M= [0,\frac{1}{2},0]$, $L=[0,\half,\half]$, and $H=[\frac{1}{3},\frac{1}{3},\half]$.
}
\label{fig:mos}
\end{figure}

\begin{figure}
\centering\includegraphics[width=9.2cm,clip]{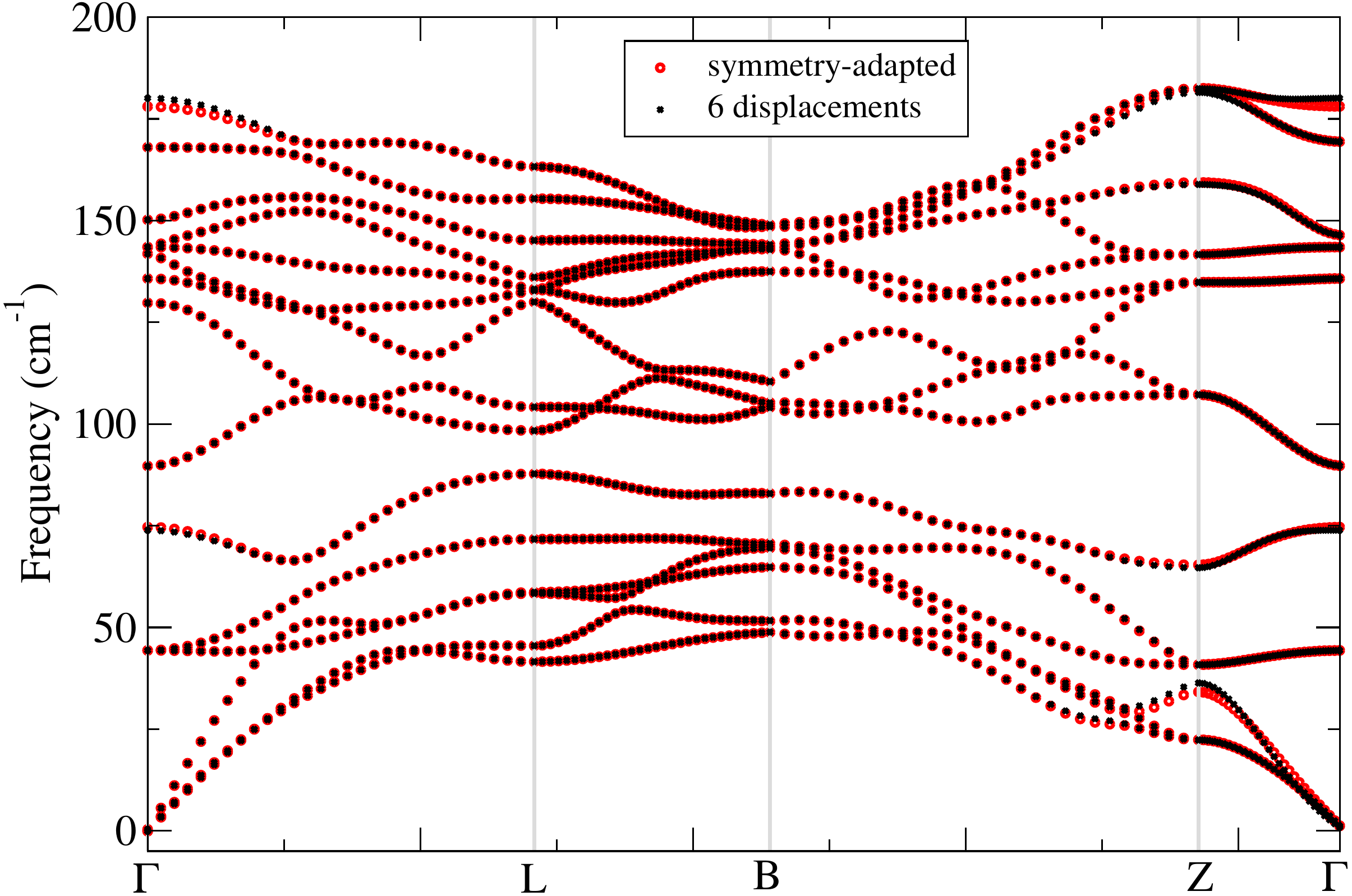}
\caption{
Phonon dispersions of trigonal \bise{} with $a_r=9.621$~\AA, 
$\alpha_r=24.64^\circ$, which corresponds to a conventional hexagonal cell of
$a= 4.105 $ and $c=27.973$~\AA{}, obtained with (a) symmetry-adapted atomic displacements and (b) six displacements for each of the three
inequivalent atoms.
The effect of longitudinal optical (LO) and transverse optical (TO) splitting\cite{Liu14v16} has been taken account.
The supercell is $4 \times 4 \times 1$ of the 
conventional hexagonal unit cell. The cutoff energy for the plane-wave basis set is $423.2$~eV. A $k$ mesh of $4 \times 4 \times 2$ is 
used for electronic relaxation.
The selected $\VEC{q}$ points (in $\VEC{b}_1$, $\VEC{b}_2$, and $\VEC{b}_3$) are 
$\Gamma=[0,0,0]$, $L= [\half,0,0]$, $B=[\eta,\half,1-\eta]$, and $Z=[\half,\half,\half]$, where $\eta = (1+4 \cos \alpha_r)/ (2+ 4\cos\alpha_r)$.\cite{Setyawan10v49}
}
\label{fig:bise}
\end{figure}

\begin{figure}
\centering\includegraphics[width=9.2cm,clip]{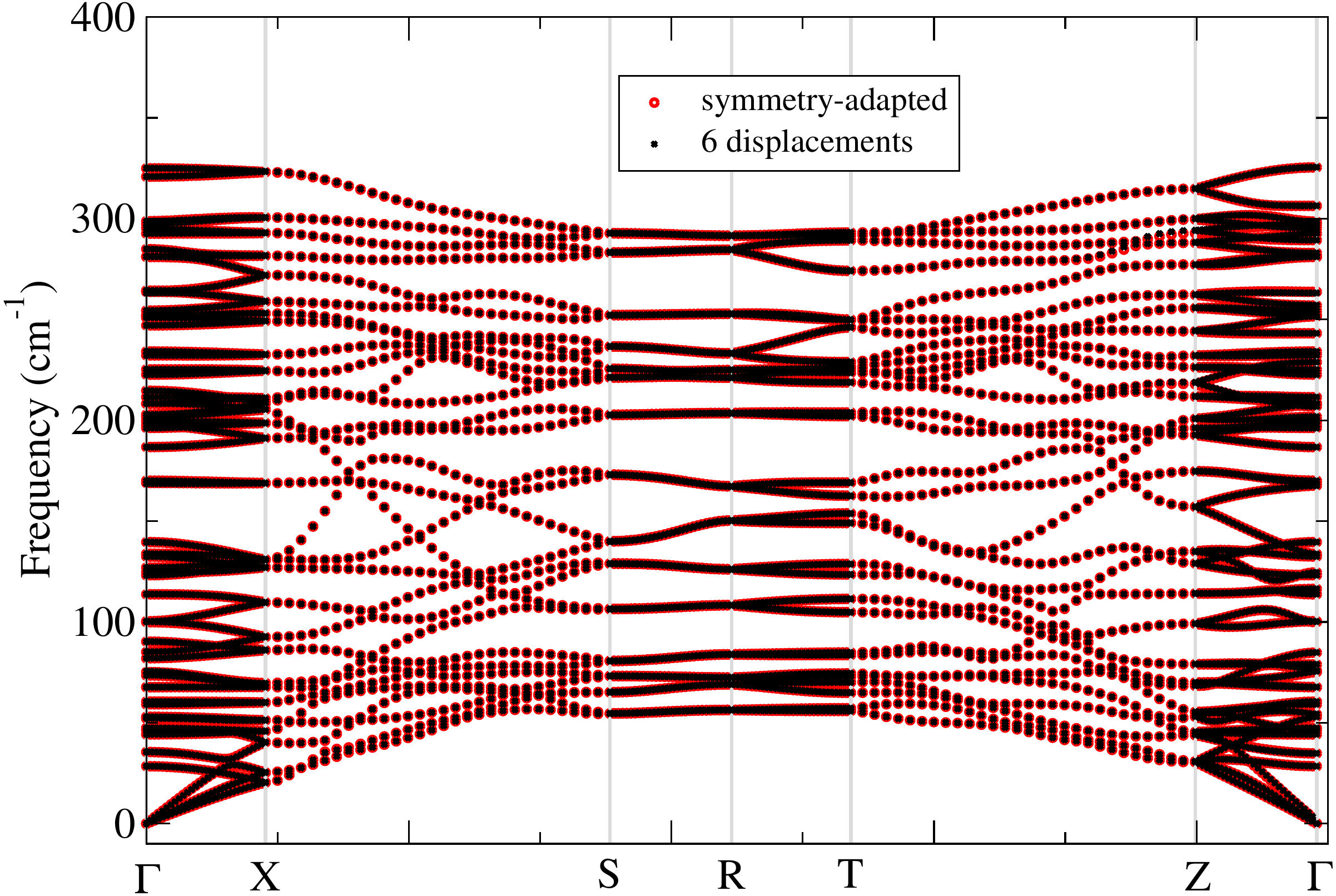}
\caption{
Phonon dispersions of orthorhombic \sbs{} with $a=11.011$, $b= 3.812 $, 
and $c=10.794$~\AA{}  obtained with (a) symmetry-adapted atomic displacements and (b) six displacements for each of the five
inequivalent atoms.
The effect of longitudinal optical (LO) and transverse optical (TO) splitting\cite{Liu14v16} has been taken account.
A $2\times 4 \times 2$ supercell is used. 
The cutoff energy for the plane-wave basis set is $323.3$~eV. A $k$ mesh of $2 \times 3 \times 2$ is 
used for electronic relaxation.
The selected $\VEC{q}$ points (in $\VEC{b}_1$, $\VEC{b}_2$, and $\VEC{b}_3$) are 
$\Gamma=[0,0,0]$, $X=[\frac{1}{2},0,0]$, $S=[\frac{1}{2},\frac{1}{2},0]$,
$R=[\frac{1}{2},\frac{1}{2},\frac{1}{2}]$,
$T=[0,\frac{1}{2},\frac{1}{2}]$, and $Z=[0,0,\frac{1}{2}]$.  
}
\label{fig:sbs}
\end{figure}

\begin{figure}
\centering\includegraphics[width=9.2cm,clip]{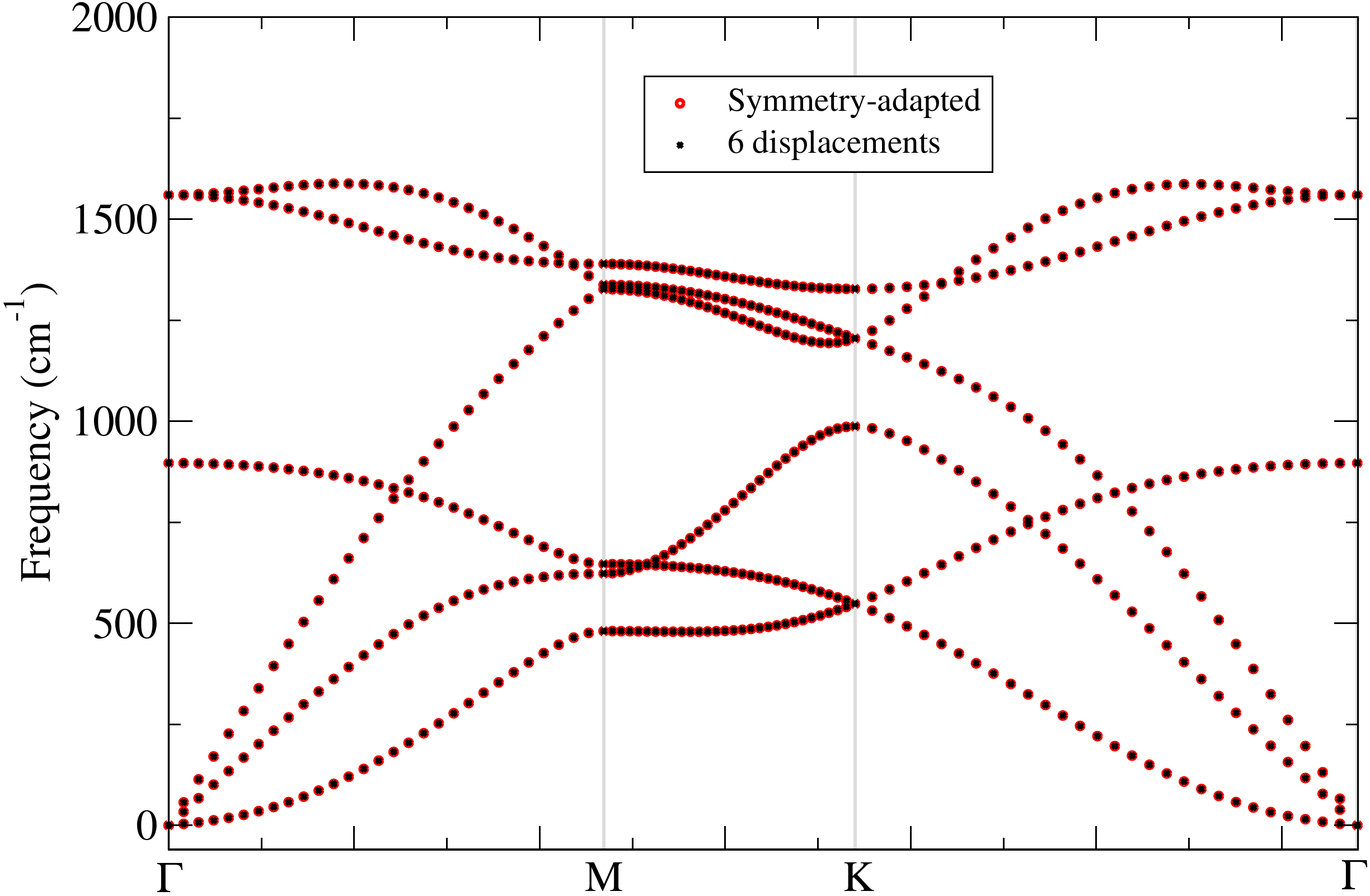}
\caption{
Phonon dispersions of 2D graphene sheet with $a=2.462$~\AA{} and a vacuum height of $12$~\AA{} obtained 
with  (a) a symmetry-adapted atomic displacement and (b) six displacements for the only
inequivalent atom.
A $4\times 4 \times 1$ supercell is used. The cutoff energy for the plane-wave basis set
is $500$~eV. A $k$ mesh of $6 \times 6 \times 1$ is 
used for electronic relaxation. 
The selected $\VEC{q}$ points (in $\VEC{b}_1$, $\VEC{b}_2$, and $\VEC{b}_3$) are 
$\Gamma=[0,0,0]$, $M= [0,\frac{1}{2},0]$, and $K = [\frac{1}{3},\frac{1}{3},0]$.
}
\label{fig:graphene}
\end{figure}

We select four representative systems to perform more phonon calculations.
For each system, we first carry out a phonon calculation by using
the symmetry-adapted atomic displacements.
Using identical computational parameters 
(i.e., energy cutoff, $k$ mesh, etc), we carry out a second phonon calculation by displacing
each inequivalent atom in the $x+$, $x-$, $y+$, $y-$, $z+$, and $z-$ along the Cartesian axes.
The results for hexagonal \mos{} [space group (SG) \# 194],  trigonal \bise{} (SG \# 166), 
orthorhombic \sbs{} (SG \# 62), and a 2D graphene sheet (SG \# 191) are shown
in Figs. \ref{fig:mos}, \ref{fig:bise}, \ref{fig:sbs}, and ~\ref{fig:graphene}, 
respectively. 
For all calculations the local density approximation is used, 
and the magnitude of the displacement $\lambda $ is $0.015$~\AA{}.
It is seen that the phonon dispersions 
obtained with both displacement methods are essentially the same that indicate
a correct implementation of the proposed methodology.
The efficacy of the symmetry-adapted atomic displacement method is shown in Table II.

\begin{table}
\begin{center}
\begin{tabular}{l|c|c|c}
\hline
  & symmetry- & $6$-  & all-  \\
  & adapted & displacement  & displacement  \\
\hline
\hline
\mos  & 3 & 12 &  36 \\
\bise & 5 & 18 & 30 \\
\sbs &  20 & 30 & 120 \\
Graphene & 1 &  6 & 12 \\
\hline
\end{tabular}
\end{center}
\caption{
The numbers of atomic displacements required
for the symmetry-adapted, 6-displacement, and all-displacement methods.
The numbers of displacements for 6-displacement and all-displacement methods are $6 n_{\rm ineq}$ and $6 n_{\rm c}$, respectively.
Here $n_{\rm ineq}$ is the number of inequivalent atoms in the unit cell and $n_{\rm c}$ is the number of atoms in the unit cell.
}
\label{tab:number}
\end{table}

\section{Conclusion}
\label{sec:conclusion}
In summary, we have proposed 
a systematic displacement method that
guarantees a theoretical optimal volume $V$
spanned by the displacement
vectors to minimize possible severe roundoff errors
for lattice dynamical studies. 
The concepts of the star of $k$ and the group of $k$ that 
are originally defined in the reciprocal space have been
extended to real space 
to facilitate a practical implementation of the method that is
readily applied to all 32 crystallographic point groups 
and all 230 space groups. 
The method generates a
minimal set of irreducible displacement vectors to
keep the number of independent force calculations to a minimum.
The error in the calculated
force constants is shown to explicitly depend on the 
inverse of the volume $V$ spanned by the displacement directions and the 
intrinsic accuracy in the induced forces.
This justified the use of the Cartesian coordinates rather 
than the fractional coordinates to optimize $V$.
Several test systems have been used to illustrate the method.
The method is shown to be very effective in dealing
cells with a large aspect ratio due to a huge difference in lattice parameters, cells with a large 
vacuum separation, and 
cells that are very oblique due to an unconventional choice of a primitive cell.
We expect the strategies employed in this paper could be extended to
deal with higher order interatomic interactions for efficiency and accuracy.

\section{Acknowledgment}

We acknowledge fruitful discussions with J. F. Kong.
We thank the National Supercomputing Center, Singapore (NSCC) and A*STAR Computational Resource Center, 
Singapore (ACRC) for computing resources. This work is
supported by RIE2020 Advanced Manufacturing and Engineering (AME) Programmatic Grant No A1898b0043.

\appendix
\section{Investigation of the determinant of $d$}

We consider a generic example of \sbs{} crystal with
space group of  $Pnma$ (SG \# $62$) which has
an orthorhombic cell.
We shall investigate the determinant of $d$ (see Eq.~\ref{eq:dmat})
as a function of cell dimensions $a$, $b$, and $c$. 
For a general discussion we
choose a primitive cell with $\VEC{a}_1 = a\VEC{i} + nb \VEC{j}$, $\VEC{a}_2 = b\VEC{j}$, and
$\VEC{a}_3 = c \VEC{k}$, where $n$ is an integer. $\VEC{i}$, $\VEC{j}$, and $\VEC{k}$ are
unit vectors in the Cartesian directions. First we 
consider $n=0$ that 
corresponds to the conventional primitive cell.

For \sbs{} crystal, all five inequivalent 
atoms occupy the $4c$ Wyckoff position hence
we may just consider any one of them. 
The site symmetry for the $4c$ position 
consists of just two elements:
\be
I = 
  \begin{pmatrix}
       1 &  0 &  0 \\
       0 &  1 &  0 \\
       0 &  0 &  1 \\
  \end{pmatrix}
\ee
and
\be
m = 
  \begin{pmatrix}
       1 &  0 &  0 \\
       0 & -1 &  0 \\
       0 &  0 &  1 \\
  \end{pmatrix}
\ee
which is a reflection across the $xz$ plane.

$\VEC{g}_1$ may be chosen to be $(1,-1,0)^T$ which will be mapped to
$\VEC{g}_2 = (1,1,0)^T$ under $m$. 
A second independent displacement is $\VEC{g}_3 =
(0,0,1)^T$.

As discussed in the main text, the actual displacement in the Cartesian coordinates is 
$ \VEC{d}_k^i =  \lambda \frac{A\VEC{g_k}}{ | A\VEC{g_k}   | }$, $k=1,2,3$. 
We find 
\bea
V &=& \det d \\
&=& \frac{ \det A} { |A\VEC{g}_1 | |A\VEC{g}_2| |A\VEC{g}_3|  } 
\\ &=&  \frac{  2 a b   }{  \sqrt{ a^2 + (n-1)^2 b^2    }   \sqrt{  a^2 + (n+1)^2 b^2   }   }
\eea
while the angle $\theta$ between 
the displacements $\VEC{d}^i_1$ and $\VEC{d}_2^i$ 
satisfies
\be
\cos \theta = \frac{  a^2 + (n^2 -1 ) b^2   }{       \sqrt{ a^2 + (n-1)^2 b^2    }   \sqrt{  a^2 + (n+1)^2 b^2   }     }
\ee

When $n=0$, we find $V = \frac{2ab}{a^2 + b^2}$ (therefore $V \rightarrow 2 (a/b)^{-1} $ for $a \gg b$)
 and $\cos \theta =
\frac{a^2 - b^2}{a^2 + b^2}$. The last result
is equivalent to a simpler expression
of $\tan^{-1} \frac{\theta}{2}  = \frac{b}{a}$, consistent
with the fact that the two displacements are $(a, -b,0)^T$ and $(a,b,0)^T$ in
the Cartesian coordinates.
If $a$ is much larger than $b$, then $V$ decreases as $\frac{2b}{a}$
to zero while $\cos \theta$ approaches $1$ since the two displacements
$\VEC{d}_1^i$ and $\VEC{d}_2^i$ 
are almost parallel.

Next we discuss the effect of $n \ne 0$. 
The site symmetry operations become\cite{Nespolo16v72} 
\be
I = 
  \begin{pmatrix}
       1 &  0 &  0 \\
       0 &  1 &  0 \\
       0 &  0 &  1 \\
  \end{pmatrix}
\ee
and
\be
m = 
  \begin{pmatrix}
       1 &  0 &  0 \\
       -2n & -1 &  0 \\
       0 &  0 &  1 \\
  \end{pmatrix}
\ee

If $\VEC{g}_1 = (1,0,0)^T$ is used, the $m$ operation on $\VEC{g}_1$ gives
$\VEC{g}_2 = (1,-2n,0)^T$. A second independent displacement is
$\VEC{g}_3 = (0,0,1)^T$.
For a simple analysis, we set $a=b$.
For large $n$, we find $V$ approaches $-\frac{2}{n} $, while
$\cos\theta $ approaches $-1$, which means the displacement $\VEC{d}_1^i$
approaches $-\VEC{d}_2^i$.

This shows that using displacement directions in the form of nonzero $(e_1,e_2,e_3)^T$
where $e_i= 0,\pm 1$, $i=1,2,3$ may not be an optimal choice.

\begin{thebibliography}{59}%
\makeatletter
\providecommand \@ifxundefined [1]{%
 \@ifx{#1\undefined}
}%
\providecommand \@ifnum [1]{%
 \ifnum #1\expandafter \@firstoftwo
 \else \expandafter \@secondoftwo
 \fi
}%
\providecommand \@ifx [1]{%
 \ifx #1\expandafter \@firstoftwo
 \else \expandafter \@secondoftwo
 \fi
}%
\providecommand \natexlab [1]{#1}%
\providecommand \enquote  [1]{``#1''}%
\providecommand \bibnamefont  [1]{#1}%
\providecommand \bibfnamefont [1]{#1}%
\providecommand \citenamefont [1]{#1}%
\providecommand \href@noop [0]{\@secondoftwo}%
\providecommand \href [0]{\begingroup \@sanitize@url \@href}%
\providecommand \@href[1]{\@@startlink{#1}\@@href}%
\providecommand \@@href[1]{\endgroup#1\@@endlink}%
\providecommand \@sanitize@url [0]{\catcode `\\12\catcode `\$12\catcode
  `\&12\catcode `\#12\catcode `\^12\catcode `\_12\catcode `\%12\relax}%
\providecommand \@@startlink[1]{}%
\providecommand \@@endlink[0]{}%
\providecommand \url  [0]{\begingroup\@sanitize@url \@url }%
\providecommand \@url [1]{\endgroup\@href {#1}{\urlprefix }}%
\providecommand \urlprefix  [0]{URL }%
\providecommand \Eprint [0]{\href }%
\providecommand \doibase [0]{https://doi.org/}%
\providecommand \selectlanguage [0]{\@gobble}%
\providecommand \bibinfo  [0]{\@secondoftwo}%
\providecommand \bibfield  [0]{\@secondoftwo}%
\providecommand \translation [1]{[#1]}%
\providecommand \BibitemOpen [0]{}%
\providecommand \bibitemStop [0]{}%
\providecommand \bibitemNoStop [0]{.\EOS\space}%
\providecommand \EOS [0]{\spacefactor3000\relax}%
\providecommand \BibitemShut  [1]{\csname bibitem#1\endcsname}%
\let\auto@bib@innerbib\@empty
\bibitem [{\citenamefont {Born}\ and\ \citenamefont
  {Huang}(1956)}]{Born56-book}%
  \BibitemOpen
  \bibfield  {author} {\bibinfo {author} {\bibfnamefont {M.}~\bibnamefont
  {Born}}\ and\ \bibinfo {author} {\bibfnamefont {K.}~\bibnamefont {Huang}},\
  }\href@noop {} {\emph {\bibinfo {title} {Dynamical Theory of Crystal
  Lattices}}}\ (\bibinfo  {publisher} {Oxford University Press},\ \bibinfo
  {address} {London},\ \bibinfo {year} {1956})\BibitemShut {NoStop}%
\bibitem [{\citenamefont {van~de Walle}\ and\ \citenamefont
  {Ceder}(2002)}]{VanDeWalle02v74}%
  \BibitemOpen
  \bibfield  {author} {\bibinfo {author} {\bibfnamefont {A.}~\bibnamefont
  {van~de Walle}}\ and\ \bibinfo {author} {\bibfnamefont {G.}~\bibnamefont
  {Ceder}},\ }\bibfield  {title} {\bibinfo {title} {The effect of lattice
  vibrations on substitutional alloy thermodynamics},\ }\href@noop {}
  {\bibfield  {journal} {\bibinfo  {journal} {Rev. Mod. Phys.}\ }\textbf
  {\bibinfo {volume} {74}},\ \bibinfo {pages} {11} (\bibinfo {year}
  {2002})}\BibitemShut {NoStop}%
\bibitem [{\citenamefont {Gan}\ \emph {et~al.}(2010)\citenamefont {Gan},
  \citenamefont {Fan},\ and\ \citenamefont {Kuo}}]{Gan10v49}%
  \BibitemOpen
  \bibfield  {author} {\bibinfo {author} {\bibfnamefont {C.~K.}\ \bibnamefont
  {Gan}}, \bibinfo {author} {\bibfnamefont {X.~F.}\ \bibnamefont {Fan}},\ and\
  \bibinfo {author} {\bibfnamefont {J.-L.}\ \bibnamefont {Kuo}},\ }\bibfield
  {title} {\bibinfo {title} {Composition-temperature phase diagram of
  {B}e$_x${Z}n$_{1-x}${O} from first principles},\ }\href@noop {} {\bibfield
  {journal} {\bibinfo  {journal} {Comput. Mater. Sci.}\ }\textbf {\bibinfo
  {volume} {49}},\ \bibinfo {pages} {S29} (\bibinfo {year} {2010})}\BibitemShut
  {NoStop}%
\bibitem [{\citenamefont {Zhong}\ \emph {et~al.}(1994)\citenamefont {Zhong},
  \citenamefont {King-Smith},\ and\ \citenamefont {Vanderbilt}}]{Zhong94v72}%
  \BibitemOpen
  \bibfield  {author} {\bibinfo {author} {\bibfnamefont {W.}~\bibnamefont
  {Zhong}}, \bibinfo {author} {\bibfnamefont {R.~D.}\ \bibnamefont
  {King-Smith}},\ and\ \bibinfo {author} {\bibfnamefont {D.}~\bibnamefont
  {Vanderbilt}},\ }\bibfield  {title} {\bibinfo {title} {Giant {LO-TO}
  splittings in perovskite ferroelectrics},\ }\href@noop {} {\bibfield
  {journal} {\bibinfo  {journal} {Phys. Rev. Lett.}\ }\textbf {\bibinfo
  {volume} {72}},\ \bibinfo {pages} {3618} (\bibinfo {year}
  {1994})}\BibitemShut {NoStop}%
\bibitem [{\citenamefont {Zhao}\ \emph {et~al.}(2013)\citenamefont {Zhao},
  \citenamefont {Luo}, \citenamefont {Li}, \citenamefont {Zhang}, \citenamefont
  {Araujo}, \citenamefont {Gan}, \citenamefont {Wu}, \citenamefont {Zhang},
  \citenamefont {Quek}, \citenamefont {Dresselhaus},\ and\ \citenamefont
  {Xiong}}]{Zhao13v13}%
  \BibitemOpen
  \bibfield  {author} {\bibinfo {author} {\bibfnamefont {Y.}~\bibnamefont
  {Zhao}}, \bibinfo {author} {\bibfnamefont {X.}~\bibnamefont {Luo}}, \bibinfo
  {author} {\bibfnamefont {H.}~\bibnamefont {Li}}, \bibinfo {author}
  {\bibfnamefont {J.}~\bibnamefont {Zhang}}, \bibinfo {author} {\bibfnamefont
  {P.~T.}\ \bibnamefont {Araujo}}, \bibinfo {author} {\bibfnamefont {C.~K.}\
  \bibnamefont {Gan}}, \bibinfo {author} {\bibfnamefont {J.}~\bibnamefont
  {Wu}}, \bibinfo {author} {\bibfnamefont {H.}~\bibnamefont {Zhang}}, \bibinfo
  {author} {\bibfnamefont {S.~Y.}\ \bibnamefont {Quek}}, \bibinfo {author}
  {\bibfnamefont {M.~S.}\ \bibnamefont {Dresselhaus}},\ and\ \bibinfo {author}
  {\bibfnamefont {Q.~H.}\ \bibnamefont {Xiong}},\ }\bibfield  {title} {\bibinfo
  {title} {Interlayer brething and shear modes in few-trilayer {M}o{S}2 and
  {W}{Se}2},\ }\href@noop {} {\bibfield  {journal} {\bibinfo  {journal} {Nano
  Lett.}\ }\textbf {\bibinfo {volume} {13}},\ \bibinfo {pages} {1007} (\bibinfo
  {year} {2013})}\BibitemShut {NoStop}%
\bibitem [{\citenamefont {Chong}\ \emph {et~al.}(2014)\citenamefont {Chong},
  \citenamefont {Xing}, \citenamefont {Liu}, \citenamefont {Gui}, \citenamefont
  {Zhang}, \citenamefont {Xiong}, \citenamefont {Mathews}, \citenamefont
  {Gan},\ and\ \citenamefont {Sum}}]{Chong14v90}%
  \BibitemOpen
  \bibfield  {author} {\bibinfo {author} {\bibfnamefont {W.~K.}\ \bibnamefont
  {Chong}}, \bibinfo {author} {\bibfnamefont {G.}~\bibnamefont {Xing}},
  \bibinfo {author} {\bibfnamefont {Y.}~\bibnamefont {Liu}}, \bibinfo {author}
  {\bibfnamefont {E.~L.}\ \bibnamefont {Gui}}, \bibinfo {author} {\bibfnamefont
  {Q.}~\bibnamefont {Zhang}}, \bibinfo {author} {\bibfnamefont
  {Q.}~\bibnamefont {Xiong}}, \bibinfo {author} {\bibfnamefont
  {N.}~\bibnamefont {Mathews}}, \bibinfo {author} {\bibfnamefont {C.~K.}\
  \bibnamefont {Gan}},\ and\ \bibinfo {author} {\bibfnamefont {T.~C.}\
  \bibnamefont {Sum}},\ }\bibfield  {title} {\bibinfo {title} {Direct
  measurement of coherent phonon dynamics in solution-processed stibnite thin
  films},\ }\href@noop {} {\bibfield  {journal} {\bibinfo  {journal} {Phys.
  Rev. B}\ }\textbf {\bibinfo {volume} {90}},\ \bibinfo {pages} {035208}
  (\bibinfo {year} {2014})}\BibitemShut {NoStop}%
\bibitem [{\citenamefont {Giovanni}\ \emph {et~al.}(2018)\citenamefont
  {Giovanni}, \citenamefont {Chong}, \citenamefont {Liu}, \citenamefont {Dewi},
  \citenamefont {Yin}, \citenamefont {Lekina}, \citenamefont {Shen},
  \citenamefont {Mathews}, \citenamefont {Gan},\ and\ \citenamefont
  {Sum}}]{Giovanni18v5}%
  \BibitemOpen
  \bibfield  {author} {\bibinfo {author} {\bibfnamefont {D.}~\bibnamefont
  {Giovanni}}, \bibinfo {author} {\bibfnamefont {W.~K.}\ \bibnamefont {Chong}},
  \bibinfo {author} {\bibfnamefont {Y.~Y.~F.}\ \bibnamefont {Liu}}, \bibinfo
  {author} {\bibfnamefont {H.~A.}\ \bibnamefont {Dewi}}, \bibinfo {author}
  {\bibfnamefont {T.}~\bibnamefont {Yin}}, \bibinfo {author} {\bibfnamefont
  {Y.}~\bibnamefont {Lekina}}, \bibinfo {author} {\bibfnamefont {Z.~X.}\
  \bibnamefont {Shen}}, \bibinfo {author} {\bibfnamefont {N.}~\bibnamefont
  {Mathews}}, \bibinfo {author} {\bibfnamefont {C.~K.}\ \bibnamefont {Gan}},\
  and\ \bibinfo {author} {\bibfnamefont {T.~C.}\ \bibnamefont {Sum}},\
  }\bibfield  {title} {\bibinfo {title} {Coherent spin and quasiparticle
  dynamics in solution-processed layered 2{D} lead halide perovskites},\
  }\href@noop {} {\bibfield  {journal} {\bibinfo  {journal} {Adv. Sci.}\
  }\textbf {\bibinfo {volume} {5}},\ \bibinfo {pages} {1800664} (\bibinfo
  {year} {2018})}\BibitemShut {NoStop}%
\bibitem [{\citenamefont {Giustino}(2017)}]{Giustino17v89}%
  \BibitemOpen
  \bibfield  {author} {\bibinfo {author} {\bibfnamefont {F.}~\bibnamefont
  {Giustino}},\ }\bibfield  {title} {\bibinfo {title} {Electron-phonon
  interactions from first-principles},\ }\href@noop {} {\bibfield  {journal}
  {\bibinfo  {journal} {Rev. Mod. Phys.}\ }\textbf {\bibinfo {volume} {89}},\
  \bibinfo {pages} {015003} (\bibinfo {year} {2017})}\BibitemShut {NoStop}%
\bibitem [{\citenamefont {Togo}\ \emph {et~al.}(2008)\citenamefont {Togo},
  \citenamefont {Oba},\ and\ \citenamefont {Tanaka}}]{Togo08v78}%
  \BibitemOpen
  \bibfield  {author} {\bibinfo {author} {\bibfnamefont {A.}~\bibnamefont
  {Togo}}, \bibinfo {author} {\bibfnamefont {F.}~\bibnamefont {Oba}},\ and\
  \bibinfo {author} {\bibfnamefont {I.}~\bibnamefont {Tanaka}},\ }\bibfield
  {title} {\bibinfo {title} {First-principles calculations of the ferroelastic
  transition between rutile-type and {C}a{C}l$_2$-type {S}i{O}$_2$ at high
  pressures},\ }\href@noop {} {\bibfield  {journal} {\bibinfo  {journal} {Phys.
  Rev. B}\ }\textbf {\bibinfo {volume} {78}},\ \bibinfo {pages} {134106}
  (\bibinfo {year} {2008})}\BibitemShut {NoStop}%
\bibitem [{\citenamefont {Grimvall}(1999)}]{Grimvall1999-book}%
  \BibitemOpen
  \bibfield  {author} {\bibinfo {author} {\bibfnamefont {G.}~\bibnamefont
  {Grimvall}},\ }\href@noop {} {\emph {\bibinfo {title} {Thermophysical
  Properties of Materials}}}\ (\bibinfo  {publisher} {Elsevier Science B.V.},\
  \bibinfo {address} {Amsterdam, The Netherlands},\ \bibinfo {year}
  {1999})\BibitemShut {NoStop}%
\bibitem [{\citenamefont {Mujica}\ \emph {et~al.}(2003)\citenamefont {Mujica},
  \citenamefont {Rubio}, \citenamefont {Munoz},\ and\ \citenamefont
  {Needs}}]{Mujica03v75}%
  \BibitemOpen
  \bibfield  {author} {\bibinfo {author} {\bibfnamefont {A.}~\bibnamefont
  {Mujica}}, \bibinfo {author} {\bibfnamefont {A.}~\bibnamefont {Rubio}},
  \bibinfo {author} {\bibfnamefont {A.}~\bibnamefont {Munoz}},\ and\ \bibinfo
  {author} {\bibfnamefont {R.~J.}\ \bibnamefont {Needs}},\ }\bibfield  {title}
  {\bibinfo {title} {High-pressure phases of group-{IV}, {III-V}, and {II-VI}
  compounds},\ }\href@noop {} {\bibfield  {journal} {\bibinfo  {journal} {Rev.
  Mod. Phys.}\ }\textbf {\bibinfo {volume} {75}},\ \bibinfo {pages} {863}
  (\bibinfo {year} {2003})}\BibitemShut {NoStop}%
\bibitem [{\citenamefont {Mounet}\ and\ \citenamefont
  {Marzari}(2005)}]{Mounet05v71}%
  \BibitemOpen
  \bibfield  {author} {\bibinfo {author} {\bibfnamefont {N.}~\bibnamefont
  {Mounet}}\ and\ \bibinfo {author} {\bibfnamefont {N.}~\bibnamefont
  {Marzari}},\ }\bibfield  {title} {\bibinfo {title} {First-principles
  determination of the structural, vibrational and thermodynamic properties of
  diamond, graphite, and derivatives},\ }\href@noop {} {\bibfield  {journal}
  {\bibinfo  {journal} {Phys. Rev. B}\ }\textbf {\bibinfo {volume} {71}},\
  \bibinfo {pages} {205214} (\bibinfo {year} {2005})}\BibitemShut {NoStop}%
\bibitem [{\citenamefont {Allen}(2020)}]{Allen20v34}%
  \BibitemOpen
  \bibfield  {author} {\bibinfo {author} {\bibfnamefont {P.~B.}\ \bibnamefont
  {Allen}},\ }\bibfield  {title} {\bibinfo {title} {Theory of thermal
  expansion: Quasi-harmonic approximation and corrections from quasi-particle
  renormalization},\ }\href@noop {} {\bibfield  {journal} {\bibinfo  {journal}
  {Mod. Phys. Lett. B}\ }\textbf {\bibinfo {volume} {34}},\ \bibinfo {pages}
  {2050025} (\bibinfo {year} {2020})}\BibitemShut {NoStop}%
\bibitem [{\citenamefont {Malica}\ and\ \citenamefont
  {Corso}(2020)}]{Malica20v32}%
  \BibitemOpen
  \bibfield  {author} {\bibinfo {author} {\bibfnamefont {C.}~\bibnamefont
  {Malica}}\ and\ \bibinfo {author} {\bibfnamefont {A.~D.}\ \bibnamefont
  {Corso}},\ }\bibfield  {title} {\bibinfo {title} {Quasi-harmonic temperature
  dependent elastic constants: Applications to silicon, aluminum, and silver},\
  }\href@noop {} {\bibfield  {journal} {\bibinfo  {journal} {J. Phys.: Condens.
  Matter}\ }\textbf {\bibinfo {volume} {32}},\ \bibinfo {pages} {315902}
  (\bibinfo {year} {2020})}\BibitemShut {NoStop}%
\bibitem [{\citenamefont {Toher}\ \emph {et~al.}(2014)\citenamefont {Toher},
  \citenamefont {Plata}, \citenamefont {Levy}, \citenamefont {de~Jong},
  \citenamefont {Asta}, \citenamefont {BuongiornoNardelli},\ and\ \citenamefont
  {Curtarolo}}]{Toher14v90}%
  \BibitemOpen
  \bibfield  {author} {\bibinfo {author} {\bibfnamefont {C.}~\bibnamefont
  {Toher}}, \bibinfo {author} {\bibfnamefont {J.~J.}\ \bibnamefont {Plata}},
  \bibinfo {author} {\bibfnamefont {O.}~\bibnamefont {Levy}}, \bibinfo {author}
  {\bibfnamefont {M.}~\bibnamefont {de~Jong}}, \bibinfo {author} {\bibfnamefont
  {M.}~\bibnamefont {Asta}}, \bibinfo {author} {\bibfnamefont {M.}~\bibnamefont
  {BuongiornoNardelli}},\ and\ \bibinfo {author} {\bibfnamefont
  {S.}~\bibnamefont {Curtarolo}},\ }\bibfield  {title} {\bibinfo {title}
  {High-throughput computational screening of thermal conductivity, {D}ebye
  temperature, and {G}r\"uneisen parameter using a quasiharmonic {D}ebye
  model},\ }\href@noop {} {\bibfield  {journal} {\bibinfo  {journal} {Phys.
  Rev. B}\ }\textbf {\bibinfo {volume} {90}},\ \bibinfo {pages} {174107}
  (\bibinfo {year} {2014})}\BibitemShut {NoStop}%
\bibitem [{\citenamefont {Snyder}\ and\ \citenamefont
  {Toberer}(2008)}]{Snyder08v7}%
  \BibitemOpen
  \bibfield  {author} {\bibinfo {author} {\bibfnamefont {G.~J.}\ \bibnamefont
  {Snyder}}\ and\ \bibinfo {author} {\bibfnamefont {E.~S.}\ \bibnamefont
  {Toberer}},\ }\bibfield  {title} {\bibinfo {title} {Complex thermoelectric
  materials},\ }\href@noop {} {\bibfield  {journal} {\bibinfo  {journal}
  {Nature Mater.}\ }\textbf {\bibinfo {volume} {7}},\ \bibinfo {pages} {105}
  (\bibinfo {year} {2008})}\BibitemShut {NoStop}%
\bibitem [{\citenamefont {Madsen}\ \emph {et~al.}(2016)\citenamefont {Madsen},
  \citenamefont {Katre},\ and\ \citenamefont {Bera}}]{Madsen16v213}%
  \BibitemOpen
  \bibfield  {author} {\bibinfo {author} {\bibfnamefont {G.~K.~H.}\
  \bibnamefont {Madsen}}, \bibinfo {author} {\bibfnamefont {A.}~\bibnamefont
  {Katre}},\ and\ \bibinfo {author} {\bibfnamefont {C.}~\bibnamefont {Bera}},\
  }\bibfield  {title} {\bibinfo {title} {Calculating the thermal conductivity
  of the silicon clathrates using the quasi-harmonic approximation},\
  }\href@noop {} {\bibfield  {journal} {\bibinfo  {journal} {Phys. Stat. Sol.
  (A)}\ }\textbf {\bibinfo {volume} {213}},\ \bibinfo {pages} {802} (\bibinfo
  {year} {2016})}\BibitemShut {NoStop}%
\bibitem [{\citenamefont {Gorai}\ \emph {et~al.}(2017)\citenamefont {Gorai},
  \citenamefont {Stevanovi\'c},\ and\ \citenamefont {Toberer}}]{Gorai17v2}%
  \BibitemOpen
  \bibfield  {author} {\bibinfo {author} {\bibfnamefont {P.}~\bibnamefont
  {Gorai}}, \bibinfo {author} {\bibfnamefont {V.}~\bibnamefont
  {Stevanovi\'c}},\ and\ \bibinfo {author} {\bibfnamefont {E.}~\bibnamefont
  {Toberer}},\ }\bibfield  {title} {\bibinfo {title} {Computationally guided
  discovery of thermoelectric materials},\ }\href@noop {} {\bibfield  {journal}
  {\bibinfo  {journal} {Nature Rev. Mat.}\ }\textbf {\bibinfo {volume} {2}},\
  \bibinfo {pages} {17053} (\bibinfo {year} {2017})}\BibitemShut {NoStop}%
\bibitem [{\citenamefont {Gan}\ \emph {et~al.}(2015)\citenamefont {Gan},
  \citenamefont {Soh},\ and\ \citenamefont {Liu}}]{Gan15v92}%
  \BibitemOpen
  \bibfield  {author} {\bibinfo {author} {\bibfnamefont {C.~K.}\ \bibnamefont
  {Gan}}, \bibinfo {author} {\bibfnamefont {J.~R.}\ \bibnamefont {Soh}},\ and\
  \bibinfo {author} {\bibfnamefont {Y.}~\bibnamefont {Liu}},\ }\bibfield
  {title} {\bibinfo {title} {Large anharmonic effect and thermal expansion
  anisotropy of metal chalcogenides: The case of antimony sulfide},\
  }\href@noop {} {\bibfield  {journal} {\bibinfo  {journal} {Phys. Rev. B}\
  }\textbf {\bibinfo {volume} {92}},\ \bibinfo {pages} {235202} (\bibinfo
  {year} {2015})}\BibitemShut {NoStop}%
\bibitem [{\citenamefont {Arnaud}\ \emph {et~al.}(2016)\citenamefont {Arnaud},
  \citenamefont {Leb\`egue},\ and\ \citenamefont {Raffy}}]{Arnaud16v93}%
  \BibitemOpen
  \bibfield  {author} {\bibinfo {author} {\bibfnamefont {B.}~\bibnamefont
  {Arnaud}}, \bibinfo {author} {\bibfnamefont {S.}~\bibnamefont {Leb\`egue}},\
  and\ \bibinfo {author} {\bibfnamefont {G.}~\bibnamefont {Raffy}},\ }\bibfield
   {title} {\bibinfo {title} {Anisotropic thermal expansion of bismuth from
  first principles},\ }\href@noop {} {\bibfield  {journal} {\bibinfo  {journal}
  {Phys. Rev. B}\ }\textbf {\bibinfo {volume} {93}},\ \bibinfo {pages} {094106}
  (\bibinfo {year} {2016})}\BibitemShut {NoStop}%
\bibitem [{\citenamefont {Gan}\ and\ \citenamefont {Liu}(2016)}]{Gan16v94}%
  \BibitemOpen
  \bibfield  {author} {\bibinfo {author} {\bibfnamefont {C.~K.}\ \bibnamefont
  {Gan}}\ and\ \bibinfo {author} {\bibfnamefont {Y.~Y.~F.}\ \bibnamefont
  {Liu}},\ }\bibfield  {title} {\bibinfo {title} {Direct calculation of the
  linear thermal expansion coefficients of {M}o{S}$_2$ via symmetry-preserving
  deformations},\ }\href@noop {} {\bibfield  {journal} {\bibinfo  {journal}
  {Phys. Rev. B}\ }\textbf {\bibinfo {volume} {94}},\ \bibinfo {pages} {134303}
  (\bibinfo {year} {2016})}\BibitemShut {NoStop}%
\bibitem [{\citenamefont {Romao}(2017)}]{Romao17v96}%
  \BibitemOpen
  \bibfield  {author} {\bibinfo {author} {\bibfnamefont {C.~P.}\ \bibnamefont
  {Romao}},\ }\bibfield  {title} {\bibinfo {title} {Anisotropic thermal
  expansion in flexible materials},\ }\href@noop {} {\bibfield  {journal}
  {\bibinfo  {journal} {Phys. Rev. B}\ }\textbf {\bibinfo {volume} {96}},\
  \bibinfo {pages} {134113} (\bibinfo {year} {2017})}\BibitemShut {NoStop}%
\bibitem [{\citenamefont {Gan}\ and\ \citenamefont {Lee}(2018)}]{Gan18v151}%
  \BibitemOpen
  \bibfield  {author} {\bibinfo {author} {\bibfnamefont {C.~K.}\ \bibnamefont
  {Gan}}\ and\ \bibinfo {author} {\bibfnamefont {C.~H.}\ \bibnamefont {Lee}},\
  }\bibfield  {title} {\bibinfo {title} {Anharmonic phonon effects on linear
  thermal expansion of trigonal bismuth selenide and antimony telluride
  crystals},\ }\href@noop {} {\bibfield  {journal} {\bibinfo  {journal}
  {Comput. Mater. Sci.}\ }\textbf {\bibinfo {volume} {151}},\ \bibinfo {pages}
  {49} (\bibinfo {year} {2018})}\BibitemShut {NoStop}%
\bibitem [{\citenamefont {Gan}\ and\ \citenamefont {Chua}(2019)}]{Gan19v31}%
  \BibitemOpen
  \bibfield  {author} {\bibinfo {author} {\bibfnamefont {C.~K.}\ \bibnamefont
  {Gan}}\ and\ \bibinfo {author} {\bibfnamefont {K.~T.~E.}\ \bibnamefont
  {Chua}},\ }\bibfield  {title} {\bibinfo {title} {Large thermal anisotropy in
  monoclinic niobium trisulfide: A thermal expansion tensor study},\
  }\href@noop {} {\bibfield  {journal} {\bibinfo  {journal} {J. Phys.: Condens.
  Matter}\ }\textbf {\bibinfo {volume} {31}},\ \bibinfo {pages} {265401}
  (\bibinfo {year} {2019})}\BibitemShut {NoStop}%
\bibitem [{\citenamefont {Lee}\ and\ \citenamefont {Gan}(2017)}]{Lee17v96}%
  \BibitemOpen
  \bibfield  {author} {\bibinfo {author} {\bibfnamefont {C.~H.}\ \bibnamefont
  {Lee}}\ and\ \bibinfo {author} {\bibfnamefont {C.~K.}\ \bibnamefont {Gan}},\
  }\bibfield  {title} {\bibinfo {title} {Anharmonic interatomic force constants
  and thermal conductivity from {G}r\"uneisen parameters: An application to
  graphene},\ }\href@noop {} {\bibfield  {journal} {\bibinfo  {journal} {Phys.
  Rev. B}\ }\textbf {\bibinfo {volume} {96}},\ \bibinfo {pages} {035105}
  (\bibinfo {year} {2017})}\BibitemShut {NoStop}%
\bibitem [{\citenamefont {Petretto}\ \emph {et~al.}(2018)\citenamefont
  {Petretto}, \citenamefont {Dwaraknath}, \citenamefont {Miranda},
  \citenamefont {Winston}, \citenamefont {Giantomassi}, \citenamefont {van
  Setten}, \citenamefont {Gonze}, \citenamefont {Persson}, \citenamefont
  {Hautier},\ and\ \citenamefont {Rignanese}}]{Petretto18v5}%
  \BibitemOpen
  \bibfield  {author} {\bibinfo {author} {\bibfnamefont {G.}~\bibnamefont
  {Petretto}}, \bibinfo {author} {\bibfnamefont {S.}~\bibnamefont
  {Dwaraknath}}, \bibinfo {author} {\bibfnamefont {H.~P.}\ \bibnamefont
  {Miranda}}, \bibinfo {author} {\bibfnamefont {D.}~\bibnamefont {Winston}},
  \bibinfo {author} {\bibfnamefont {M.}~\bibnamefont {Giantomassi}}, \bibinfo
  {author} {\bibfnamefont {M.~J.}\ \bibnamefont {van Setten}}, \bibinfo
  {author} {\bibfnamefont {X.}~\bibnamefont {Gonze}}, \bibinfo {author}
  {\bibfnamefont {K.~A.}\ \bibnamefont {Persson}}, \bibinfo {author}
  {\bibfnamefont {G.}~\bibnamefont {Hautier}},\ and\ \bibinfo {author}
  {\bibfnamefont {G.-M.}\ \bibnamefont {Rignanese}},\ }\bibfield  {title}
  {\bibinfo {title} {Data descriptor: High-throughput density-functional
  perturbation theory phonons for inorganic materials},\ }\href@noop {}
  {\bibfield  {journal} {\bibinfo  {journal} {Scientific Data}\ }\textbf
  {\bibinfo {volume} {5}},\ \bibinfo {pages} {180065} (\bibinfo {year}
  {2018})}\BibitemShut {NoStop}%
\bibitem [{\citenamefont {Togo}(2020{\natexlab{a}})}]{TogoPhononDB2020-link}%
  \BibitemOpen
  \bibfield  {author} {\bibinfo {author} {\bibfnamefont {A.}~\bibnamefont
  {Togo}},\ }\href@noop {} {\bibinfo {title} {Phonon database at {K}yoto
  {U}niversity}},\ \bibinfo {howpublished}
  {\url{http://phonondb.mtl.kyoto-u.ac.jp}} (\bibinfo {year}
  {2020}{\natexlab{a}})\BibitemShut {NoStop}%
\bibitem [{\citenamefont {Curtarolo}\ \emph {et~al.}(2013)\citenamefont
  {Curtarolo}, \citenamefont {Hart}, \citenamefont {BuongiornoNardelli},
  \citenamefont {Mingo}, \citenamefont {Sanvito},\ and\ \citenamefont
  {Levy}}]{Curtarolo13v12}%
  \BibitemOpen
  \bibfield  {author} {\bibinfo {author} {\bibfnamefont {S.}~\bibnamefont
  {Curtarolo}}, \bibinfo {author} {\bibfnamefont {G.~L.~W.}\ \bibnamefont
  {Hart}}, \bibinfo {author} {\bibfnamefont {M.}~\bibnamefont
  {BuongiornoNardelli}}, \bibinfo {author} {\bibfnamefont {N.}~\bibnamefont
  {Mingo}}, \bibinfo {author} {\bibfnamefont {S.}~\bibnamefont {Sanvito}},\
  and\ \bibinfo {author} {\bibfnamefont {O.}~\bibnamefont {Levy}},\ }\bibfield
  {title} {\bibinfo {title} {The high-throughput highway to computational
  materials design},\ }\href@noop {} {\bibfield  {journal} {\bibinfo  {journal}
  {Nature Mater.}\ }\textbf {\bibinfo {volume} {12}},\ \bibinfo {pages} {191}
  (\bibinfo {year} {2013})}\BibitemShut {NoStop}%
\bibitem [{\citenamefont {Ong}\ \emph {et~al.}(2013)\citenamefont {Ong},
  \citenamefont {Richards}, \citenamefont {Jain}, \citenamefont {Hautier},
  \citenamefont {Kocher}, \citenamefont {Cholia}, \citenamefont {Gunter},
  \citenamefont {Chevrier}, \citenamefont {Persson},\ and\ \citenamefont
  {Ceder}}]{Ong13v68}%
  \BibitemOpen
  \bibfield  {author} {\bibinfo {author} {\bibfnamefont {S.~P.}\ \bibnamefont
  {Ong}}, \bibinfo {author} {\bibfnamefont {W.~D.}\ \bibnamefont {Richards}},
  \bibinfo {author} {\bibfnamefont {A.}~\bibnamefont {Jain}}, \bibinfo {author}
  {\bibfnamefont {G.}~\bibnamefont {Hautier}}, \bibinfo {author} {\bibfnamefont
  {M.}~\bibnamefont {Kocher}}, \bibinfo {author} {\bibfnamefont
  {S.}~\bibnamefont {Cholia}}, \bibinfo {author} {\bibfnamefont
  {D.}~\bibnamefont {Gunter}}, \bibinfo {author} {\bibfnamefont {V.~L.}\
  \bibnamefont {Chevrier}}, \bibinfo {author} {\bibfnamefont {K.~A.}\
  \bibnamefont {Persson}},\ and\ \bibinfo {author} {\bibfnamefont
  {G.}~\bibnamefont {Ceder}},\ }\bibfield  {title} {\bibinfo {title} {Python
  materials genomics (pymatgen): A robus, open-source python library for
  material analysis},\ }\href@noop {} {\bibfield  {journal} {\bibinfo
  {journal} {Comput. Mater. Sci.}\ }\textbf {\bibinfo {volume} {68}},\ \bibinfo
  {pages} {314} (\bibinfo {year} {2013})}\BibitemShut {NoStop}%
\bibitem [{\citenamefont {Pizzi}\ \emph {et~al.}(2016)\citenamefont {Pizzi},
  \citenamefont {Cepellotti}, \citenamefont {Sabatini}, \citenamefont
  {Marzari},\ and\ \citenamefont {Kozinsky}}]{Pizzi16v111}%
  \BibitemOpen
  \bibfield  {author} {\bibinfo {author} {\bibfnamefont {G.}~\bibnamefont
  {Pizzi}}, \bibinfo {author} {\bibfnamefont {A.}~\bibnamefont {Cepellotti}},
  \bibinfo {author} {\bibfnamefont {R.}~\bibnamefont {Sabatini}}, \bibinfo
  {author} {\bibfnamefont {N.}~\bibnamefont {Marzari}},\ and\ \bibinfo {author}
  {\bibfnamefont {B.}~\bibnamefont {Kozinsky}},\ }\bibfield  {title} {\bibinfo
  {title} {Aiida: automated interactive infrastructure and database for
  computational science},\ }\href@noop {} {\bibfield  {journal} {\bibinfo
  {journal} {Comput. Mater. Sci.}\ }\textbf {\bibinfo {volume} {111}},\
  \bibinfo {pages} {218} (\bibinfo {year} {2016})}\BibitemShut {NoStop}%
\bibitem [{\citenamefont {Parlinski}\ \emph {et~al.}(1997)\citenamefont
  {Parlinski}, \citenamefont {Li},\ and\ \citenamefont
  {Kawazoe}}]{Parlinski97v78}%
  \BibitemOpen
  \bibfield  {author} {\bibinfo {author} {\bibfnamefont {K.}~\bibnamefont
  {Parlinski}}, \bibinfo {author} {\bibfnamefont {Z.~Q.}\ \bibnamefont {Li}},\
  and\ \bibinfo {author} {\bibfnamefont {Y.}~\bibnamefont {Kawazoe}},\
  }\bibfield  {title} {\bibinfo {title} {First-principles determination of the
  soft mode in cubic {Z}r{O}$_2$},\ }\href@noop {} {\bibfield  {journal}
  {\bibinfo  {journal} {Phys. Rev. Lett.}\ }\textbf {\bibinfo {volume} {78}},\
  \bibinfo {pages} {4063} (\bibinfo {year} {1997})}\BibitemShut {NoStop}%
\bibitem [{\citenamefont {Wang}\ \emph {et~al.}(2014)\citenamefont {Wang},
  \citenamefont {Chen},\ and\ \citenamefont {Liu}}]{Wang14v185}%
  \BibitemOpen
  \bibfield  {author} {\bibinfo {author} {\bibfnamefont {Y.}~\bibnamefont
  {Wang}}, \bibinfo {author} {\bibfnamefont {L.-Q.}\ \bibnamefont {Chen}},\
  and\ \bibinfo {author} {\bibfnamefont {Z.-K.}\ \bibnamefont {Liu}},\
  }\bibfield  {title} {\bibinfo {title} {{YPHON}: A package for calculating
  phonons of polar materials},\ }\href@noop {} {\bibfield  {journal} {\bibinfo
  {journal} {Comput. Phys. Comm.}\ }\textbf {\bibinfo {volume} {185}},\
  \bibinfo {pages} {2950} (\bibinfo {year} {2014})}\BibitemShut {NoStop}%
\bibitem [{\citenamefont {Wang}\ \emph {et~al.}(2016)\citenamefont {Wang},
  \citenamefont {Shang}, \citenamefont {Fang}, \citenamefont {Liu},\ and\
  \citenamefont {Chen}}]{Wang16v2}%
  \BibitemOpen
  \bibfield  {author} {\bibinfo {author} {\bibfnamefont {Y.}~\bibnamefont
  {Wang}}, \bibinfo {author} {\bibfnamefont {S.-L.}\ \bibnamefont {Shang}},
  \bibinfo {author} {\bibfnamefont {H.}~\bibnamefont {Fang}}, \bibinfo {author}
  {\bibfnamefont {Z.-K.}\ \bibnamefont {Liu}},\ and\ \bibinfo {author}
  {\bibfnamefont {L.-Q.}\ \bibnamefont {Chen}},\ }\bibfield  {title} {\bibinfo
  {title} {First-principles calculations of lattice dynamics and thermal
  properties of polar solids},\ }\href@noop {} {\bibfield  {journal} {\bibinfo
  {journal} {npj Comp. Mater.}\ }\textbf {\bibinfo {volume} {2}},\ \bibinfo
  {pages} {16006} (\bibinfo {year} {2016})}\BibitemShut {NoStop}%
\bibitem [{\citenamefont {Ackland}\ \emph {et~al.}(1997)\citenamefont
  {Ackland}, \citenamefont {Warren},\ and\ \citenamefont
  {Clark}}]{Ackland97v9}%
  \BibitemOpen
  \bibfield  {author} {\bibinfo {author} {\bibfnamefont {G.~J.}\ \bibnamefont
  {Ackland}}, \bibinfo {author} {\bibfnamefont {M.~C.}\ \bibnamefont
  {Warren}},\ and\ \bibinfo {author} {\bibfnamefont {S.~J.}\ \bibnamefont
  {Clark}},\ }\bibfield  {title} {\bibinfo {title} {Practical methods in ab
  initio lattice dynamics},\ }\href@noop {} {\bibfield  {journal} {\bibinfo
  {journal} {J. Phys.: Condens. Matter}\ }\textbf {\bibinfo {volume} {9}},\
  \bibinfo {pages} {7861} (\bibinfo {year} {1997})}\BibitemShut {NoStop}%
\bibitem [{\citenamefont {Togo}\ and\ \citenamefont
  {Tanaka}(2015)}]{Togo15v108}%
  \BibitemOpen
  \bibfield  {author} {\bibinfo {author} {\bibfnamefont {A.}~\bibnamefont
  {Togo}}\ and\ \bibinfo {author} {\bibfnamefont {I.}~\bibnamefont {Tanaka}},\
  }\bibfield  {title} {\bibinfo {title} {First principles phonon calculations
  in materials science},\ }\href@noop {} {\bibfield  {journal} {\bibinfo
  {journal} {Scr. Mater.}\ }\textbf {\bibinfo {volume} {108}},\ \bibinfo
  {pages} {1} (\bibinfo {year} {2015})}\BibitemShut {NoStop}%
\bibitem [{\citenamefont {Togo}(2020{\natexlab{b}})}]{Togo2020-github}%
  \BibitemOpen
  \bibfield  {author} {\bibinfo {author} {\bibfnamefont {A.}~\bibnamefont
  {Togo}},\ }\href@noop {} {\bibinfo {title} {Phonopy}},\ \bibinfo
  {howpublished} {\url{https://phonopy.github.io/phonopy}} (\bibinfo {year}
  {2020}{\natexlab{b}})\BibitemShut {NoStop}%
\bibitem [{\citenamefont {Alf\`e}(2009)}]{Alfe09v180}%
  \BibitemOpen
  \bibfield  {author} {\bibinfo {author} {\bibfnamefont {D.}~\bibnamefont
  {Alf\`e}},\ }\bibfield  {title} {\bibinfo {title} {Phon: A program to
  calculate phonons using the small displacement method},\ }\href@noop {}
  {\bibfield  {journal} {\bibinfo  {journal} {Comput. Phys. Comm.}\ }\textbf
  {\bibinfo {volume} {180}},\ \bibinfo {pages} {2622} (\bibinfo {year}
  {2009})}\BibitemShut {NoStop}%
\bibitem [{\citenamefont {Kresse}\ \emph {et~al.}(1995)\citenamefont {Kresse},
  \citenamefont {Furthm\"uller},\ and\ \citenamefont {Hafner}}]{Kresse95v32}%
  \BibitemOpen
  \bibfield  {author} {\bibinfo {author} {\bibfnamefont {G.}~\bibnamefont
  {Kresse}}, \bibinfo {author} {\bibfnamefont {J.}~\bibnamefont
  {Furthm\"uller}},\ and\ \bibinfo {author} {\bibfnamefont {J.}~\bibnamefont
  {Hafner}},\ }\bibfield  {title} {\bibinfo {title} {Ab initio force constant
  approach to phonon dispersion relations of diamond and graphite},\
  }\href@noop {} {\bibfield  {journal} {\bibinfo  {journal} {Europhys. Lett.}\
  }\textbf {\bibinfo {volume} {32}},\ \bibinfo {pages} {729} (\bibinfo {year}
  {1995})}\BibitemShut {NoStop}%
\bibitem [{\citenamefont {Feynman}(1939)}]{Feynman39v56}%
  \BibitemOpen
  \bibfield  {author} {\bibinfo {author} {\bibfnamefont {R.~P.}\ \bibnamefont
  {Feynman}},\ }\bibfield  {title} {\bibinfo {title} {Forces in molecules},\
  }\href@noop {} {\bibfield  {journal} {\bibinfo  {journal} {Phys. Rev.}\
  }\textbf {\bibinfo {volume} {56}},\ \bibinfo {pages} {340} (\bibinfo {year}
  {1939})}\BibitemShut {NoStop}%
\bibitem [{\citenamefont {Lloyd-Williams}\ and\ \citenamefont
  {Monserrat}(2015)}]{Lloyd-Williams15v92}%
  \BibitemOpen
  \bibfield  {author} {\bibinfo {author} {\bibfnamefont {J.~H.}\ \bibnamefont
  {Lloyd-Williams}}\ and\ \bibinfo {author} {\bibfnamefont {B.}~\bibnamefont
  {Monserrat}},\ }\bibfield  {title} {\bibinfo {title} {Lattice dynamics and
  electron-phonon coupling calculations using nondiagonal supercells},\
  }\href@noop {} {\bibfield  {journal} {\bibinfo  {journal} {Phys. Rev. B}\
  }\textbf {\bibinfo {volume} {92}},\ \bibinfo {pages} {184301} (\bibinfo
  {year} {2015})}\BibitemShut {NoStop}%
\bibitem [{\citenamefont {Baroni}\ \emph {et~al.}(2001)\citenamefont {Baroni},
  \citenamefont {{de Gironcoli}}, \citenamefont {{Dal Corso}},\ and\
  \citenamefont {Giannozzi}}]{Baroni01v73}%
  \BibitemOpen
  \bibfield  {author} {\bibinfo {author} {\bibfnamefont {S.}~\bibnamefont
  {Baroni}}, \bibinfo {author} {\bibfnamefont {S.}~\bibnamefont {{de
  Gironcoli}}}, \bibinfo {author} {\bibfnamefont {A.}~\bibnamefont {{Dal
  Corso}}},\ and\ \bibinfo {author} {\bibfnamefont {P.}~\bibnamefont
  {Giannozzi}},\ }\bibfield  {title} {\bibinfo {title} {Phonons and related
  crystal properties from density-functional perturbation theory},\ }\href@noop
  {} {\bibfield  {journal} {\bibinfo  {journal} {Rev. Mod. Phys.}\ }\textbf
  {\bibinfo {volume} {73}},\ \bibinfo {pages} {515} (\bibinfo {year}
  {2001})}\BibitemShut {NoStop}%
\bibitem [{\citenamefont {Gonze}(1997)}]{Gonze97v55a}%
  \BibitemOpen
  \bibfield  {author} {\bibinfo {author} {\bibfnamefont {X.}~\bibnamefont
  {Gonze}},\ }\bibfield  {title} {\bibinfo {title} {First-principles responses
  of solids to atomic displacements and homogeneous electric fields:
  Implementation of a conjugate-gradient algorithm},\ }\href@noop {} {\bibfield
   {journal} {\bibinfo  {journal} {Phys. Rev. B}\ }\textbf {\bibinfo {volume}
  {55}},\ \bibinfo {pages} {10337} (\bibinfo {year} {1997})}\BibitemShut
  {NoStop}%
\bibitem [{\citenamefont {Giannozzi}\ \emph {et~al.}(2009)\citenamefont
  {Giannozzi}, \citenamefont {Baroni}, \citenamefont {Bonini}, \citenamefont
  {Calandra}, \citenamefont {Car}, \citenamefont {Cavazzoni}, \citenamefont
  {Ceresoli}, \citenamefont {Chiarotti}, \citenamefont {Cococcioni},
  \citenamefont {Dabo}, \citenamefont {{Dal Corso}}, \citenamefont {{de
  Gironcoli}}, \citenamefont {Fabris}, \citenamefont {Fratesi}, \citenamefont
  {Gebauer}, \citenamefont {Gerstmann}, \citenamefont {Gougoussis},
  \citenamefont {Kokalj}, \citenamefont {Lazzeri}, \citenamefont
  {Martin-Samos}, \citenamefont {Marzari}, \citenamefont {Mauri}, \citenamefont
  {Mazzarello}, \citenamefont {Paolini}, \citenamefont {Pasquarello},
  \citenamefont {Paulatto}, \citenamefont {Sbraccia}, \citenamefont {Scandolo},
  \citenamefont {Sclauzero}, \citenamefont {Seitsonen}, \citenamefont
  {Smogunov}, \citenamefont {Umari},\ and\ \citenamefont
  {Wentzcovitch}}]{Giannozzi09v21}%
  \BibitemOpen
  \bibfield  {author} {\bibinfo {author} {\bibfnamefont {P.}~\bibnamefont
  {Giannozzi}}, \bibinfo {author} {\bibfnamefont {S.}~\bibnamefont {Baroni}},
  \bibinfo {author} {\bibfnamefont {N.}~\bibnamefont {Bonini}}, \bibinfo
  {author} {\bibfnamefont {M.}~\bibnamefont {Calandra}}, \bibinfo {author}
  {\bibfnamefont {R.}~\bibnamefont {Car}}, \bibinfo {author} {\bibfnamefont
  {C.}~\bibnamefont {Cavazzoni}}, \bibinfo {author} {\bibfnamefont
  {D.}~\bibnamefont {Ceresoli}}, \bibinfo {author} {\bibfnamefont {G.~L.}\
  \bibnamefont {Chiarotti}}, \bibinfo {author} {\bibfnamefont {M.}~\bibnamefont
  {Cococcioni}}, \bibinfo {author} {\bibfnamefont {I.}~\bibnamefont {Dabo}},
  \bibinfo {author} {\bibfnamefont {A.}~\bibnamefont {{Dal Corso}}}, \bibinfo
  {author} {\bibfnamefont {S.}~\bibnamefont {{de Gironcoli}}}, \bibinfo
  {author} {\bibfnamefont {S.}~\bibnamefont {Fabris}}, \bibinfo {author}
  {\bibfnamefont {G.}~\bibnamefont {Fratesi}}, \bibinfo {author} {\bibfnamefont
  {R.}~\bibnamefont {Gebauer}}, \bibinfo {author} {\bibfnamefont
  {U.}~\bibnamefont {Gerstmann}}, \bibinfo {author} {\bibfnamefont
  {C.}~\bibnamefont {Gougoussis}}, \bibinfo {author} {\bibfnamefont
  {A.}~\bibnamefont {Kokalj}}, \bibinfo {author} {\bibfnamefont
  {M.}~\bibnamefont {Lazzeri}}, \bibinfo {author} {\bibfnamefont
  {L.}~\bibnamefont {Martin-Samos}}, \bibinfo {author} {\bibfnamefont
  {N.}~\bibnamefont {Marzari}}, \bibinfo {author} {\bibfnamefont
  {F.}~\bibnamefont {Mauri}}, \bibinfo {author} {\bibfnamefont
  {R.}~\bibnamefont {Mazzarello}}, \bibinfo {author} {\bibfnamefont
  {S.}~\bibnamefont {Paolini}}, \bibinfo {author} {\bibfnamefont
  {A.}~\bibnamefont {Pasquarello}}, \bibinfo {author} {\bibfnamefont
  {L.}~\bibnamefont {Paulatto}}, \bibinfo {author} {\bibfnamefont
  {C.}~\bibnamefont {Sbraccia}}, \bibinfo {author} {\bibfnamefont
  {S.}~\bibnamefont {Scandolo}}, \bibinfo {author} {\bibfnamefont
  {G.}~\bibnamefont {Sclauzero}}, \bibinfo {author} {\bibfnamefont {A.~P.}\
  \bibnamefont {Seitsonen}}, \bibinfo {author} {\bibfnamefont {A.}~\bibnamefont
  {Smogunov}}, \bibinfo {author} {\bibfnamefont {P.}~\bibnamefont {Umari}},\
  and\ \bibinfo {author} {\bibfnamefont {R.~M.}\ \bibnamefont {Wentzcovitch}},\
  }\bibfield  {title} {\bibinfo {title} {{Q}uantum {E}spresso: a modular and
  open-source software project for quantum simulations of materials},\
  }\href@noop {} {\bibfield  {journal} {\bibinfo  {journal} {J. Phys.: Condens.
  Matter}\ }\textbf {\bibinfo {volume} {21}},\ \bibinfo {pages} {395502}
  (\bibinfo {year} {2009})}\BibitemShut {NoStop}%
\bibitem [{\citenamefont {Gonze}\ \emph {et~al.}(2009)\citenamefont {Gonze},
  \citenamefont {Amadon}, \citenamefont {Anglade}, \citenamefont {Beuken},
  \citenamefont {Bottin}, \citenamefont {Boulanger}, \citenamefont {Bruneval},
  \citenamefont {Caliste}, \citenamefont {Caracas}, \citenamefont {Cote},
  \citenamefont {Deutsch}, \citenamefont {Genovese}, \citenamefont {Ghosez},
  \citenamefont {Giantomassi}, \citenamefont {Goedecker}, \citenamefont
  {Hamann}, \citenamefont {Hermet}, \citenamefont {Jollet}, \citenamefont
  {Jomard}, \citenamefont {Leroux}, \citenamefont {Mancini}, \citenamefont
  {Mazevet}, \citenamefont {Oliveir}, \citenamefont {Onidab}, \citenamefont
  {Pouillon}, \citenamefont {Rangel}, \citenamefont {Rignanese}, \citenamefont
  {Sangalli}, \citenamefont {Shaltaf}, \citenamefont {Torrent}, \citenamefont
  {Verstraete}, \citenamefont {Zerah},\ and\ \citenamefont
  {Zwanziger}}]{Gonze09v180}%
  \BibitemOpen
  \bibfield  {author} {\bibinfo {author} {\bibfnamefont {X.}~\bibnamefont
  {Gonze}}, \bibinfo {author} {\bibfnamefont {B.}~\bibnamefont {Amadon}},
  \bibinfo {author} {\bibfnamefont {P.-M.}\ \bibnamefont {Anglade}}, \bibinfo
  {author} {\bibfnamefont {J.-M.}\ \bibnamefont {Beuken}}, \bibinfo {author}
  {\bibfnamefont {F.}~\bibnamefont {Bottin}}, \bibinfo {author} {\bibfnamefont
  {P.}~\bibnamefont {Boulanger}}, \bibinfo {author} {\bibfnamefont
  {F.}~\bibnamefont {Bruneval}}, \bibinfo {author} {\bibfnamefont
  {D.}~\bibnamefont {Caliste}}, \bibinfo {author} {\bibfnamefont
  {R.}~\bibnamefont {Caracas}}, \bibinfo {author} {\bibfnamefont
  {M.}~\bibnamefont {Cote}}, \bibinfo {author} {\bibfnamefont {T.}~\bibnamefont
  {Deutsch}}, \bibinfo {author} {\bibfnamefont {L.}~\bibnamefont {Genovese}},
  \bibinfo {author} {\bibfnamefont {P.}~\bibnamefont {Ghosez}}, \bibinfo
  {author} {\bibfnamefont {M.}~\bibnamefont {Giantomassi}}, \bibinfo {author}
  {\bibfnamefont {S.}~\bibnamefont {Goedecker}}, \bibinfo {author}
  {\bibfnamefont {D.}~\bibnamefont {Hamann}}, \bibinfo {author} {\bibfnamefont
  {P.}~\bibnamefont {Hermet}}, \bibinfo {author} {\bibfnamefont
  {F.}~\bibnamefont {Jollet}}, \bibinfo {author} {\bibfnamefont
  {G.}~\bibnamefont {Jomard}}, \bibinfo {author} {\bibfnamefont
  {S.}~\bibnamefont {Leroux}}, \bibinfo {author} {\bibfnamefont
  {M.}~\bibnamefont {Mancini}}, \bibinfo {author} {\bibfnamefont
  {S.}~\bibnamefont {Mazevet}}, \bibinfo {author} {\bibfnamefont
  {M.}~\bibnamefont {Oliveir}}, \bibinfo {author} {\bibfnamefont
  {G.}~\bibnamefont {Onidab}}, \bibinfo {author} {\bibfnamefont
  {Y.}~\bibnamefont {Pouillon}}, \bibinfo {author} {\bibfnamefont
  {T.}~\bibnamefont {Rangel}}, \bibinfo {author} {\bibfnamefont {G.-M.}\
  \bibnamefont {Rignanese}}, \bibinfo {author} {\bibfnamefont {D.}~\bibnamefont
  {Sangalli}}, \bibinfo {author} {\bibfnamefont {R.}~\bibnamefont {Shaltaf}},
  \bibinfo {author} {\bibfnamefont {M.}~\bibnamefont {Torrent}}, \bibinfo
  {author} {\bibfnamefont {M.}~\bibnamefont {Verstraete}}, \bibinfo {author}
  {\bibfnamefont {G.}~\bibnamefont {Zerah}},\ and\ \bibinfo {author}
  {\bibfnamefont {J.}~\bibnamefont {Zwanziger}},\ }\bibfield  {title} {\bibinfo
  {title} {{ABINIT}: First-principles approach to material and nanosystem
  properties},\ }\href@noop {} {\bibfield  {journal} {\bibinfo  {journal}
  {Comput. Phys. Comm.}\ }\textbf {\bibinfo {volume} {180}},\ \bibinfo {pages}
  {2582} (\bibinfo {year} {2009})}\BibitemShut {NoStop}%
\bibitem [{\citenamefont {Fu}\ \emph {et~al.}(2019)\citenamefont {Fu},
  \citenamefont {Kornbluth}, \citenamefont {Cheng},\ and\ \citenamefont
  {Marianetti}}]{Fu19v100}%
  \BibitemOpen
  \bibfield  {author} {\bibinfo {author} {\bibfnamefont {L.}~\bibnamefont
  {Fu}}, \bibinfo {author} {\bibfnamefont {M.}~\bibnamefont {Kornbluth}},
  \bibinfo {author} {\bibfnamefont {Z.}~\bibnamefont {Cheng}},\ and\ \bibinfo
  {author} {\bibfnamefont {C.~A.}\ \bibnamefont {Marianetti}},\ }\bibfield
  {title} {\bibinfo {title} {Group theoretical approach to computing phonons
  and their interactions},\ }\href@noop {} {\bibfield  {journal} {\bibinfo
  {journal} {Phys. Rev. B}\ }\textbf {\bibinfo {volume} {100}},\ \bibinfo
  {pages} {014303} (\bibinfo {year} {2019})}\BibitemShut {NoStop}%
\bibitem [{\citenamefont {Parlinski}(2018)}]{Parlinski18v98}%
  \BibitemOpen
  \bibfield  {author} {\bibinfo {author} {\bibfnamefont {K.}~\bibnamefont
  {Parlinski}},\ }\bibfield  {title} {\bibinfo {title} {Ab initio determination
  of anharmonic phonon peaks},\ }\href@noop {} {\bibfield  {journal} {\bibinfo
  {journal} {Phys. Rev. B}\ }\textbf {\bibinfo {volume} {98}},\ \bibinfo
  {pages} {054305} (\bibinfo {year} {2018})}\BibitemShut {NoStop}%
\bibitem [{\citenamefont {Hahn(Ed.)}(2006)}]{ITtable06-book}%
  \BibitemOpen
  \bibfield  {author} {\bibinfo {author} {\bibfnamefont {T.}~\bibnamefont
  {Hahn(Ed.)}},\ }\href@noop {} {\emph {\bibinfo {title} {International Tables
  for Crystallography (2006). Vol. A, Space-group symmetry}}}\ (\bibinfo
  {publisher} {Chester},\ \bibinfo {address} {International Union of
  Crystallography},\ \bibinfo {year} {2006})\BibitemShut {NoStop}%
\bibitem [{\citenamefont {Dresselhaus}\ \emph {et~al.}(2008)\citenamefont
  {Dresselhaus}, \citenamefont {Dresselhaus},\ and\ \citenamefont
  {Jorio}}]{Dresselhaus2008-book}%
  \BibitemOpen
  \bibfield  {author} {\bibinfo {author} {\bibfnamefont {M.~S.}\ \bibnamefont
  {Dresselhaus}}, \bibinfo {author} {\bibfnamefont {G.}~\bibnamefont
  {Dresselhaus}},\ and\ \bibinfo {author} {\bibfnamefont {A.}~\bibnamefont
  {Jorio}},\ }\href@noop {} {\emph {\bibinfo {title} {Group theory: Application
  to the physics of condensed matter}}}\ (\bibinfo  {publisher}
  {Springer-Verlag},\ \bibinfo {address} {Berlin, Heidelberg},\ \bibinfo {year}
  {2008})\BibitemShut {NoStop}%
\bibitem [{\citenamefont {Burns}(1985)}]{Burns85-book}%
  \BibitemOpen
  \bibfield  {author} {\bibinfo {author} {\bibfnamefont {G.}~\bibnamefont
  {Burns}},\ }\href@noop {} {\emph {\bibinfo {title} {Solid State Physics}}}\
  (\bibinfo  {publisher} {Academic Press},\ \bibinfo {address} {Orlando,
  Florida},\ \bibinfo {year} {1985})\BibitemShut {NoStop}%
\bibitem [{\citenamefont {Liu}\ \emph {et~al.}(2014)\citenamefont {Liu},
  \citenamefont {Chua}, \citenamefont {Sum},\ and\ \citenamefont
  {Gan}}]{Liu14v16}%
  \BibitemOpen
  \bibfield  {author} {\bibinfo {author} {\bibfnamefont {Y.}~\bibnamefont
  {Liu}}, \bibinfo {author} {\bibfnamefont {K.~T.~E.}\ \bibnamefont {Chua}},
  \bibinfo {author} {\bibfnamefont {T.~C.}\ \bibnamefont {Sum}},\ and\ \bibinfo
  {author} {\bibfnamefont {C.~K.}\ \bibnamefont {Gan}},\ }\bibfield  {title}
  {\bibinfo {title} {First-principles study of the lattice dynamics of
  {S}b$_2${S}$_3$},\ }\href@noop {} {\bibfield  {journal} {\bibinfo  {journal}
  {Phys. Chem. Chem. Phys.}\ }\textbf {\bibinfo {volume} {16}},\ \bibinfo
  {pages} {345} (\bibinfo {year} {2014})}\BibitemShut {NoStop}%
\bibitem [{\citenamefont {Payne}\ \emph {et~al.}(1992)\citenamefont {Payne},
  \citenamefont {Teter}, \citenamefont {Allan}, \citenamefont {Arias},\ and\
  \citenamefont {Joannopoulos}}]{Payne92v64}%
  \BibitemOpen
  \bibfield  {author} {\bibinfo {author} {\bibfnamefont {M.~C.}\ \bibnamefont
  {Payne}}, \bibinfo {author} {\bibfnamefont {M.~P.}\ \bibnamefont {Teter}},
  \bibinfo {author} {\bibfnamefont {D.~C.}\ \bibnamefont {Allan}}, \bibinfo
  {author} {\bibfnamefont {T.~A.}\ \bibnamefont {Arias}},\ and\ \bibinfo
  {author} {\bibfnamefont {J.~D.}\ \bibnamefont {Joannopoulos}},\ }\bibfield
  {title} {\bibinfo {title} {Iterative minimization techniques for ab initio
  total-energy calculations: molecular dynamics and conjugate gradients},\
  }\href@noop {} {\bibfield  {journal} {\bibinfo  {journal} {Rev. Mod. Phys.}\
  }\textbf {\bibinfo {volume} {64}},\ \bibinfo {pages} {1045} (\bibinfo {year}
  {1992})}\BibitemShut {NoStop}%
\bibitem [{\citenamefont {Gan}\ \emph {et~al.}(2001)\citenamefont {Gan},
  \citenamefont {Haynes},\ and\ \citenamefont {Payne}}]{Gan01v134}%
  \BibitemOpen
  \bibfield  {author} {\bibinfo {author} {\bibfnamefont {C.~K.}\ \bibnamefont
  {Gan}}, \bibinfo {author} {\bibfnamefont {P.~D.}\ \bibnamefont {Haynes}},\
  and\ \bibinfo {author} {\bibfnamefont {M.~C.}\ \bibnamefont {Payne}},\
  }\bibfield  {title} {\bibinfo {title} {Preconditioned conjugate gradient
  method for the sparse generalized eigenvalue problem in electronic structure
  calculations},\ }\href@noop {} {\bibfield  {journal} {\bibinfo  {journal}
  {Comput. Phys. Comm.}\ }\textbf {\bibinfo {volume} {134}},\ \bibinfo {pages}
  {33} (\bibinfo {year} {2001})}\BibitemShut {NoStop}%
\bibitem [{\citenamefont {Altmann}(1991)}]{Altmann-book}%
  \BibitemOpen
  \bibfield  {author} {\bibinfo {author} {\bibfnamefont {S.~L.}\ \bibnamefont
  {Altmann}},\ }\href@noop {} {\emph {\bibinfo {title} {Band theory of solids:
  An introduction from the point of view of symmetry}}}\ (\bibinfo  {publisher}
  {Oxford University Press},\ \bibinfo {address} {Walton Street, Oxford OX2
  6DP},\ \bibinfo {year} {1991})\BibitemShut {NoStop}%
\bibitem [{\citenamefont {Zhao}\ \emph {et~al.}(2011)\citenamefont {Zhao},
  \citenamefont {Chua}, \citenamefont {Gan}, \citenamefont {Zhang},
  \citenamefont {Peng}, \citenamefont {Peng},\ and\ \citenamefont
  {Xiong}}]{Zhao11v84}%
  \BibitemOpen
  \bibfield  {author} {\bibinfo {author} {\bibfnamefont {Y.~Y.}\ \bibnamefont
  {Zhao}}, \bibinfo {author} {\bibfnamefont {K.~T.~E.}\ \bibnamefont {Chua}},
  \bibinfo {author} {\bibfnamefont {C.~K.}\ \bibnamefont {Gan}}, \bibinfo
  {author} {\bibfnamefont {J.}~\bibnamefont {Zhang}}, \bibinfo {author}
  {\bibfnamefont {B.}~\bibnamefont {Peng}}, \bibinfo {author} {\bibfnamefont
  {Z.~P.}\ \bibnamefont {Peng}},\ and\ \bibinfo {author} {\bibfnamefont
  {Q.~H.}\ \bibnamefont {Xiong}},\ }\bibfield  {title} {\bibinfo {title}
  {Phonons in {B}i$_2${S}$_3$ nanostructures: Raman scattering and
  first-principles studies},\ }\href@noop {} {\bibfield  {journal} {\bibinfo
  {journal} {Phys. Rev. B}\ }\textbf {\bibinfo {volume} {84}},\ \bibinfo
  {pages} {205330} (\bibinfo {year} {2011})}\BibitemShut {NoStop}%
\bibitem [{\citenamefont {Gan}\ and\ \citenamefont
  {Srolovitz}(2010)}]{Gan10v81}%
  \BibitemOpen
  \bibfield  {author} {\bibinfo {author} {\bibfnamefont {C.~K.}\ \bibnamefont
  {Gan}}\ and\ \bibinfo {author} {\bibfnamefont {D.~J.}\ \bibnamefont
  {Srolovitz}},\ }\bibfield  {title} {\bibinfo {title} {First-principles study
  of graphene edge properties and flake shapes},\ }\href@noop {} {\bibfield
  {journal} {\bibinfo  {journal} {Phys. Rev. B}\ }\textbf {\bibinfo {volume}
  {81}},\ \bibinfo {pages} {125445} (\bibinfo {year} {2010})}\BibitemShut
  {NoStop}%
\bibitem [{Note1()}]{Note1}%
  \BibitemOpen
  \bibinfo {note} {See an implementation of our algorithm in fm-forces.f90 from
  https://github.com/qphonon/atomic-displacement}\BibitemShut {NoStop}%
\bibitem [{\citenamefont {Kresse}\ and\ \citenamefont
  {Furthm{\"u}ller}(1996)}]{Kresse96v6}%
  \BibitemOpen
  \bibfield  {author} {\bibinfo {author} {\bibfnamefont {G.}~\bibnamefont
  {Kresse}}\ and\ \bibinfo {author} {\bibfnamefont {J.}~\bibnamefont
  {Furthm{\"u}ller}},\ }\bibfield  {title} {\bibinfo {title} {Efficiency of
  ab-initio total energy calculations for metals and semiconductors using a
  plane-wave basis set},\ }\href@noop {} {\bibfield  {journal} {\bibinfo
  {journal} {Comput. Mater. Sci.}\ }\textbf {\bibinfo {volume} {6}},\ \bibinfo
  {pages} {15} (\bibinfo {year} {1996})}\BibitemShut {NoStop}%
\bibitem [{\citenamefont {Setyawan}\ and\ \citenamefont
  {Curtarolo}(2010)}]{Setyawan10v49}%
  \BibitemOpen
  \bibfield  {author} {\bibinfo {author} {\bibfnamefont {W.}~\bibnamefont
  {Setyawan}}\ and\ \bibinfo {author} {\bibfnamefont {S.}~\bibnamefont
  {Curtarolo}},\ }\bibfield  {title} {\bibinfo {title} {High-throughput
  electronic band structure calculations: Challenges and tools},\ }\href@noop
  {} {\bibfield  {journal} {\bibinfo  {journal} {Comput. Mater. Sci.}\ }\textbf
  {\bibinfo {volume} {49}},\ \bibinfo {pages} {299} (\bibinfo {year}
  {2010})}\BibitemShut {NoStop}%
\bibitem [{\citenamefont {Nespolo}\ and\ \citenamefont
  {Aroyo}(2016)}]{Nespolo16v72}%
  \BibitemOpen
  \bibfield  {author} {\bibinfo {author} {\bibfnamefont {M.}~\bibnamefont
  {Nespolo}}\ and\ \bibinfo {author} {\bibfnamefont {M.~I.}\ \bibnamefont
  {Aroyo}},\ }\bibfield  {title} {\bibinfo {title} {The crystallographic
  chameleon: when space groups change skin},\ }\href@noop {} {\bibfield
  {journal} {\bibinfo  {journal} {Acta Cryst. A}\ }\textbf {\bibinfo {volume}
  {72}},\ \bibinfo {pages} {523} (\bibinfo {year} {2016})}\BibitemShut
  {NoStop}%
\end{thebibliography}
\end{document}